\documentclass[journal]{IEEEtran}

\ifCLASSINFOpdf
\else
\fi

\usepackage{setspace}
\usepackage{bbm}
\usepackage[cmex10]{amsmath}
\usepackage{amssymb}
\usepackage{cite}
\usepackage{graphicx}
\usepackage{array,color}
\usepackage{amsmath}
\allowdisplaybreaks
\usepackage{stfloats}
\usepackage{graphicx}
\usepackage{epstopdf}
\usepackage{subfigure}
\usepackage{tabularx}
\usepackage{epsfig,epsf,color,balance,cite}
\usepackage{algorithmic}
\usepackage{algorithm}
\usepackage{url}
\usepackage{bm}
\usepackage{multirow}

\usepackage{amsthm}

\ifodd 0
\usepackage{soul}
\usepackage{color}
\setstcolor{red}

\newcommand{\rev}[1]{{\color{red}#1}} 
\newcommand{\del}[1]{\st{#1}} 

\newcommand{\com}[1]{\textbf{\color{red} (COMMENT: #1)}} 
\newcommand{\response}[1]{\textbf{\color{green} (RESPONSE: #1)}} 
\else

\newcommand{\rev}[1]{#1}

\newcommand{\del}[1]{}

\newcommand{\com}[1]{}
\newcommand{\comg}[1]{}
\newcommand{\response}[1]{}
\fi

\title{\huge {Intelligent Reflecting Surface-Aided Multiuser Communication: Co-design of Transmit Diversity and Active/Passive Precoding}}

\author{ 
	Beixiong Zheng,~\IEEEmembership{Senior Member,~IEEE}, Tiantian Ma, Jie Tang,~\IEEEmembership{Senior Member,~IEEE}, Changsheng You, \\
	Shaoe Lin, and Kai-Kit Wong,~\IEEEmembership{Fellow,~IEEE}
	
	\thanks{
		B. Zheng, T. Ma, and J. Tang are with the School of Microelectronics/School of Electronic and Information Engineering, South China University of Technology, Guangzhou 511442, China (e-mail: bxzheng@scut.edu.cn; mitiantianma@mail.scut.edu.cn; eejtang@scut.edu.cn).
		
		C. You  is with the Department of Electronic and Electrical Engineering, Southern University of Science and Technology (SUSTech),
		Shenzhen 518055, China (e-mail: youcs@sustech.edu.cn).
		
		S. Lin is with the School of Information Science and Technology, Guangdong University of Foreign Studies, Guangzhou 510006, China (e-mail: linshaoe@oamail.gdufs.edu.cn).
		
		K. K. Wong is with the Department of Electronic and Electrical Engineering, University College London, London, WC1E 6BT, U.K. (e-mail: kai-kit.wong@ucl.ac.uk).

	}
}

\begin{document}
\maketitle
\begin{abstract}
Intelligent reflecting surface (IRS) has become a cost-effective solution for constructing a smart and adaptive radio environment.
Most previous works on IRS have jointly designed the active and passive precoding based on perfectly or partially known channel state information (CSI).
However, in delay-sensitive or high-mobility communications, it is imperative to explore more effective methods for leveraging IRS to enhance communication reliability without the need for any CSI.
In this paper, we investigate an innovative IRS-aided multiuser communication system, which integrates an IRS with its aided multi-antenna base station (BS) to simultaneously serve multiple high-mobility users through transmit diversity and multiple low-mobility users through active/passive precoding.
In specific, we first reveal that when dynamically tuning the IRS's common phase-shift shared with all reflecting elements,
its passive precoding gain to any low-mobility user remains unchanged.
Inspired by this property, we utilize the design of common phase-shift at the IRS for achieving transmit diversity to serve high-mobility users, yet without requiring any CSI at the BS. Meanwhile, the active/passive precoding design is incorporated into the IRS-integrated BS to serve low-mobility users (assuming the CSI is known). 
Then, taking into account the interference among different users, we formulate and solve a joint optimization problem of the IRS's reflect precoding and the BS's transmit precoding, with the aim of minimizing the total transmit power at the BS.
Simulation results demonstrate that our proposed co-design of transmit diversity and active/passive precoding in IRS-aided multiuser systems can achieve superior and desirable performance compared to other benchmarks.
	
\end{abstract}
\IEEEpeerreviewmaketitle

\section{Introduction}
Thanks to recent advances in digitally-controlled metasurfaces, intelligent reflecting surface (IRS), or its equivalents such as reconfigurable intelligent surface (RIS),
has come forth as a promising technology capable of cost-effectively constructing the ``smart radio environment" \cite{wu2020intelligent,zheng2021survey,qingqing2019towards,Renzo2019Smart,Huang2020Holographic,basar2019wireless}.
This is attributed to its capacity of dynamically altering the wireless propagation environment and/or steering electromagnetic waves into desirable directions through controllable passive signal reflection, which essentially differs from traditional transceiver techniques that either combat or adapt to the largely random wireless channels. 
Specifically, IRS is an electromagnetic metasurface made up of numerous passive elements, each capable of being independently adjusted with ultra-low power consumption to tune the amplitude and/or phase-shift of the incident signal \cite{Huang2019Reconfigurable,Wu2019TWC}. 
As a result, by dynamically configuring its massive reflecting elements according to system requirements, IRS is able to create favorable channel conditions and various functions, such as enhancing desired signal power, suppressing undesired interference, and refining channel statistics \cite{wu2020intelligent}. 
Moreover, as compared to traditional active arrays equipped at the base station (BS) or relay, IRS only passively reflects the radio signals while requiring no radio-frequency (RF) chains to process/amplify signals, thus significantly reducing power consumption and hardware cost.
Not only conceptually appealing, IRS can also be practically fabricated with lightweight, low profile, and conformal geometry, thus enabling its adaptable and dense deployment in future wireless communication networks to boost both spectral and energy efficiency \cite{wu2020intelligent,zheng2021survey,qingqing2019towards,Renzo2019Smart}.
Those appealing advantages of IRS have led to extensive investigation into its application across various communication systems, such as relaying communication \cite{zheng2021irs,Yildirim2021Hybrid,Abdullah2021Optimization}, orthogonal frequency division multiplexing (OFDM) \cite{zheng2019intelligent,zheng2020intelligent,yang2019intelligent,Zheng2020Fast}, and multiple access \cite{Zheng2020IRSNOMA,Guo2021Intelligent,Zuo2021Reconfigurable},
among others.

In the literature, prior works on IRS-aided wireless communication have primarily focused on jointly designing the active and passive precoding for directional signal enhancement under the assumption of perfectly or partially known channel state information (CSI) \cite{wu2020intelligent,zheng2021survey}. 
For example, given the full CSI, the joint optimization problem of active and passive precoding has been extensively investigated in single/double/multi-IRS systems \cite{Zheng2020DoubleIRS,zheng2020efficient,mei2021intelligent,huang2021Multi-Hop,zheng2022intelligent,you2019progressive}, multi-antenna communications \cite{Mishra2024Transmitter,huang2020reconfigurable,Jinglian2023Reconfigurable}, and multi-cell networks \cite{Pan2020Multicell,Xie2021Max,Luo2021Reconfigurable,lyu2021hybrid}.
Note that achieving finer-grained passive precoding gains with IRS generally requires more CSI and/or beam training, which, however, can be practically challenging to acquire and maintain in dynamic communication environment. 
Furthermore, since IRS typically consists of numerous passive elements that lack transmission and processing functionalities, the conventional ``all-at-once" IRS channel estimation and beam training schemes (please refer to
e.g., \cite{zheng2019intelligent,zheng2020intelligent,yang2019intelligent,Zheng2020Fast,You2020Fast,wei2021channel,he2019cascaded}) will encounter long training overhead that scales proportionally with the number of reflecting elements, leading to an inevitable long delay before data transmission.
While such delay might be affordable in quasi-static or low-mobility communication scenarios where the training overhead is still shorter than the channel coherence interval, it is unacceptable in delay-sensitive or high-mobility communication scenarios \cite{Chen2022Robust,Sun2021channel,huang2021transforming,Al-Hilo2022Reconfigurable}.
In addition, the fast time-varying channel may cause beam misalignment for IRS's passive precoding, further degrading the transmission performance and posing more challenges in addressing the CSI/beam tracking problem.
This severely limits IRS's effectiveness in delay-sensitive or high-mobility communication scenarios, where the CSI may quickly become outdated or even unavailable. 
As such, besides IRS's passive precoding for low-mobility communications,
it is essential to explore innovative IRS-aided transmission schemes for high-mobility communications to improve communication reliability with very little or even no CSI.


\begin{figure}[!t]
	\centering
	\includegraphics[width=3.5in]{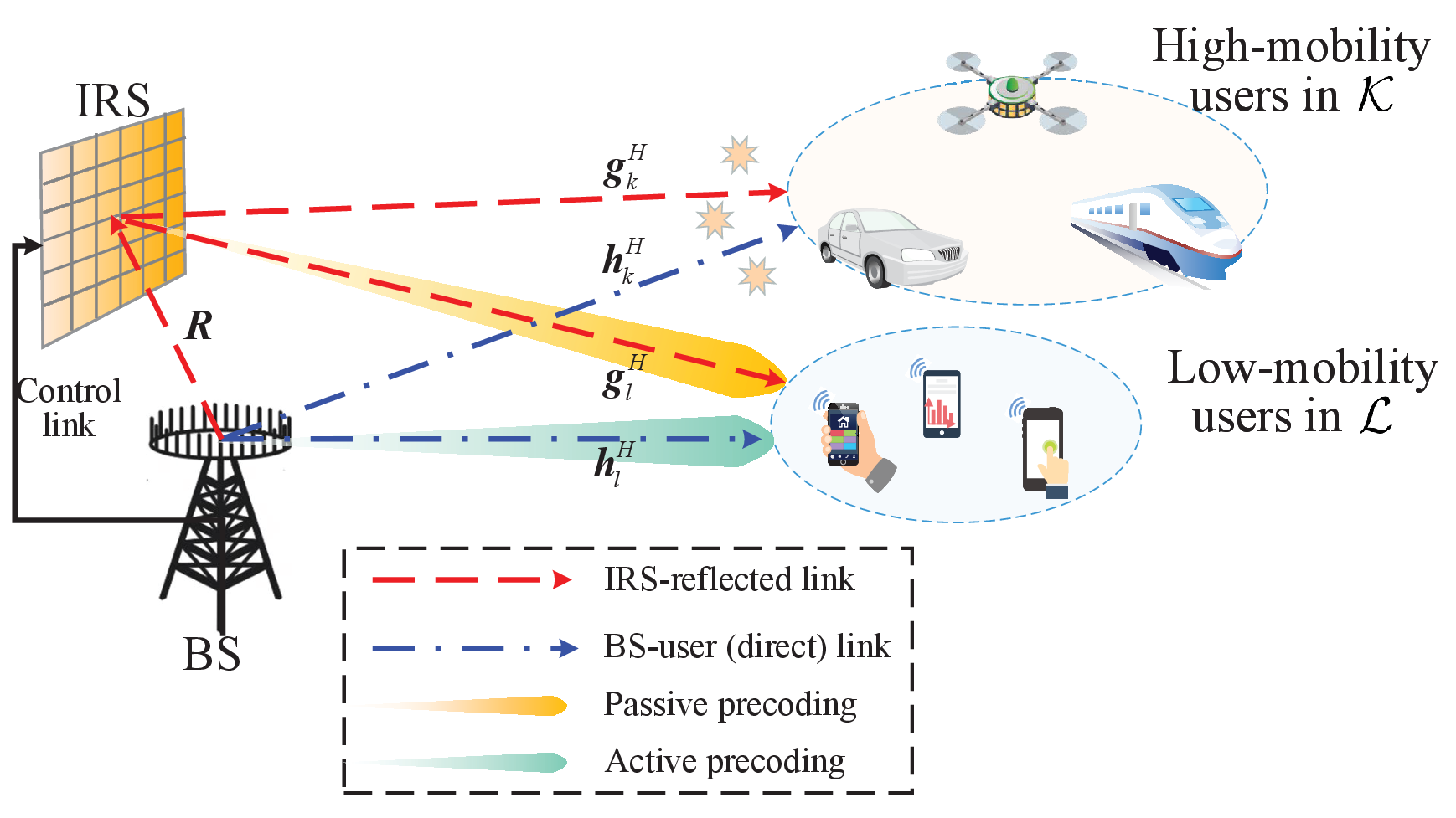}
	\setlength{\abovecaptionskip}{-3pt}
	\caption{IRS-aided multiuser downlink transmission for simultaneously serving high-mobility and low-mobility users over one given frequency band.}
	\label{system}
\end{figure}
Inspired by the considerations above, this paper explores an innovative IRS-aided multiuser downlink communication system, as depicted in Fig.~\ref{system}.
In this system, a multi-antenna BS, aided by an IRS, simultaneously serves multiple high-mobility and low-mobility users without and with CSI, respectively.
Additionally, we consider a new transmission architecture by taking the IRS as an integral part of the BS, which not only enables real-time adjustment of IRS reflection by the BS, but also minimizes the BS-IRS distance to mitigate the cascaded path loss in reflected links via IRS for all users.
Based on this architecture, we introduce a novel function of IRS to achieve  transmit diversity at the BS (without requiring users' CSI), along with its traditional function of passive precoding (generally with users' full CSI required).         
Under the above setup, we develop a practical communication protocol to serve high-mobility and low-mobility users simultaneously via transmit diversity and active/passive precoding, respectively.\footnote{Note that the system and problem in this paper differ from those in our previous work \cite{Zheng2022Simultaneous} where a single-antenna BS serves two users over two orthogonal frequency bands (without multi-user interference) and the asymptotic performance for IRS is analyzed.}
For the co-existence of high-mobility and low-mobility users with different communication requirements and/or channel conditions over one given frequency band, how to effectively cope with their mutual interference becomes a critical challenge, especially with the commonly shared IRS.
Attentive to this, we collaboratively design the IRS's transmit diversity and passive precoding additional to the BS's active precoding to minimize the
overall transmit power at the BS while adhering to individual signal-to-interference-plus-noise ratio (SINR) constraints for all users.
The primary contributions of this paper are outlined in the following.

\begin{itemize}
	\item First, we develop an IRS-aided transmit diversity scheme for serving multiple high-mobility users 
	without the need of any CSI, by carefully designing a new type of space-time code at the IRS-integrated BS.
	Specifically, given the active precoding at the multi-antenna BS, the ``passive" IRS rotates its common phase-shift in real time according to the phase difference between two modulation symbols generated by the BS for high-mobility users.
	In this way, the BS's active signal transmission and IRS's common phase rotation co-create an orthogonal channel condition to achieve transmit diversity for high-mobility users.
	\item Second, we incorporate the conventional active/passive precoding at the IRS-integrated BS to serve multiple low-mobility users simultaneously assuming that their CSI is known. This is accomplished by utilizing a compelling fact that the passive precoding gain of IRS remains unchanged to any low-mobility user when dynamically tuning IRS's common phase-shift to achieve transmit diversity.
	\item Third, we formulate a new problem to minimize the total transmit power at the BS for serving multiple high-mobility and low-mobility users simultaneously via transmit diversity and active/passive precoding, respectively, subject to their
	individual SINR constraints. This problem, however, is difficult to solve due to the coupled active/passive precoding as well as the mutual interference among users. To tackle this issue, we develop an effective algorithm leveraging the alternating optimization (AO) of the IRS's reflect precoding and the BS's transmit precoding, in an iterative manner. Specifically, the reflect precoding optimization is solved via the semidefinite relaxation (SDR); while the transmit precoding optimization can be solved via either
	the minimum mean squared error (MMSE) by virtue of the uplink-downlink duality or zero-forcing (ZF) by employing the null space of the effective channel matrix associated with low-mobility users.
	\item Finally, we provide substantial numerical results to verify the superior performance of our proposed IRS-aided multiuser communication system compared to other baseline systems, by virtue of the new architecture of the IRS-integrated BS.
	Besides, extensive simulation results further validate the effectiveness of our proposed co-design of transmit diversity and active/passive precoding.
\end{itemize}

The rest of this paper is organized as follows.
In Section~\ref{sys}, we present the system model and communication protocol for the IRS-aided multiuser downlink communication.
In Section~\ref{Model}, we introduce the      novel co-design of transmit diversity and active/passive precoding for simultaneously serving high-mobility and low-mobility users, respectively, as well as the corresponding problem formulation.
Efficient algorithms are proposed in Section~\ref{Solution} to solve the formulated problem.
In Section~\ref{Sim}, we provide extensive simulation results for evaluating the performance of proposed designs.
Finally, we draw some conclusions in Section~\ref{conlusion}.

\emph{Notation:}
Column vectors and matrices are denoted by lower-case and upper-case boldface letters, respectively.
Discrete and finite sets are denoted by upper-case calligraphic letters (e.g., $\cal{L}$).
Superscripts ${\left(\cdot\right)}^{T}$, ${\left(\cdot\right)}^{H}$, ${\left(\cdot\right)}^{-1}$ and ${\left(\cdot\right)}^{*}$ represent the transpose, Hermitian transpose, matrix inversion and conjugate operations, respectively.
${\mathbb C}^{a\times b}$ stands for the space of ${a\times b}$ complex-valued matrices.
$|\cdot|$ returns the absolute value if applied to a complex-valued number or the cardinality if applied to a set.
For a complex-valued vector $\bm{x}$,
$\angle (\bm{x} )$ denotes the phase of each element in $\bm{x}$,
$\lVert\bm{x}\rVert$ stands for its $\ell_2 $-norm,
and
$\mathrm{diag} (\bm{x})$ represents a diagonal matrix with all elements in $\bm{x}$ on its main diagonal. 
The standard big-O notation is denoted as ${\cal O}(\cdot)$,
${\mathbb P}\left\{ \cdot \right\}$ stands for the normalized principal eigenvector of a matrix,
and ${\mathbb E}\{\cdot\}$ denotes the statistical expectation.
For a complex-valued matrix ${\bm S}$, ${\rm rank} \left( {\bm S} \right)$ returns its rank,
${\rm tr}({\bm S})$ returns its trace,
 $[{\bm S}]_{i,j}$ returns its $(i,j)$-th entry,
and ${\bm S} \succeq 0$ suggests that it is positive semi-definite.
${\bm 0}$, ${\bm 1}$, and ${\bm I}$ stand for an all-zero vector/matrix, an all-one vector/matrix, and an identity matrix, respectively, with suitable dimensions.
${\mathcal N_c }({\bm \mu}, {\bm \Sigma} )$ represents the distribution of a circularly symmetric complex Gaussian (CSCG) random vector with mean vector $\bm \mu$ and covariance matrix ${\bm \Sigma}$;
and $\sim$ represents ``distributed as".

\section{System Model and Communication Protocol}\label{sys}

\subsection{System Model}
We consider a downlink communication system where a multi-antenna BS and an auxiliary IRS work collaboratively to serve multiple single-antenna users, as depicted in Fig. \ref{system}.
Notably, we assume that the IRS comprising $N$ reflecting elements is an integral part of the BS equipped with $M$ transmit antennas, thus enabling real-time signaling control and information exchange from the BS to IRS.
This new architecture, termed as the ``IRS-integrated BS" throughout this paper, enables the BS to replace the traditional IRS controller and directly control IRS for real-time reflection adjustment.

Depending on their mobility, we simply classify users into two groups: low-mobility and high-mobility users, represented by the sets ${\cal{L}}=\{1,\ldots, L\}$ and ${\cal{K}}=\{ L+1,\ldots, L+K\}$, respectively, where the corresponding set sizes, $L$ and $K$, stand for the numbers of users within each group.
Note that due to different mobility, users in the two groups generally exhibit distinct channel conditions and communication requirements. Attentive to this, we propose two communication modes at the IRS-integrated BS, namely ``transmit diversity" and ``active/passive precoding" for serving the high-mobility and low-mobility users simultaneously, without and with CSI, respectively.
In addition, we consider two basic multiuser downlink communication models for the two groups: 1) downlink multicasting (MC), where the BS sends common information to high-mobility users in ${\cal{K}}$; 2) downlink broadcasting (BC), where the BS sends independent information to different low-mobility users in ${\cal{L}}$.\footnote{If the BS needs to deliver independent information to each high-mobility user (i.e., downlink BC),  time division multiple access (TDMA) can be applied for serving different users in ${\cal{K}}$ over different time slots. }

Let ${\bm R} \in {\mathbb{C}^{N\times M }}$, ${\bm h}_{j}^H\in {\mathbb{C}^{1\times M}}$, and ${\bm g}_{j}^H\in {\mathbb{C}^{1\times N}}$ stand for the baseband equivalent channels for the BS$\rightarrow$IRS, BS$\rightarrow$user $j$, and IRS$\rightarrow$user $j$ links, respectively, with $j\in{\cal{L}}\cup{\cal{K}}$.
\rev{For high-mobility users in ${\cal{K}}$, considering the long distance and rich scattering environment between them and the BS/IRS (see Fig. \ref{system}), we adopt the Rayleigh fading channel models for both the BS$\rightarrow$user $k$ channel ${\bm h}_{k}^H$ and IRS$\rightarrow$user $k$ channel ${\bm g}_{k}^H$, which are given by 
\begin{align}\label{Rayleigh}
{\bm h}_{k}\sim {\mathcal N_c }\left({\bm 0}, \frac{\beta}{d_{k}^\alpha} {\bm I}_M \right),  {\bm g}_{k}\sim {\mathcal N_c } \left({\bm 0}, \frac{2\beta}{d_{k}^\alpha} {\bm I}_N\right), \forall k\in {\cal{K}}
\end{align}
where $\beta$ represents the reference path gain at the distance of 1 meter (m), while
$\alpha$ and $d_{k}$ denote the path loss exponent and propagation distance between the IRS-integrated BS and user~$k$, respectively.
Note that the factor of $2$ in the latter covariance accounts for the half-space reflection of each IRS element.
On the other hand, the BS$\rightarrow$IRS channel ${\bm R}$ and other channels associated with low-mobility users in ${\cal{L}}$, i.e., $\left\{{\bm g}_{l}^H, {\bm h}_{l}^H\right\}, \forall l\in{\cal{L}}$, can be completely arbitrary without assuming any specific channel model in this paper.}

On the other hand, we denote the IRS reflection vector as ${\bm \theta}\triangleq[{\theta}_{1},{\theta}_{2},\ldots,{\theta}_{N}]^T$.
For the ease of hardware implementation as well as maximizing the signal reflection power, we set the reflection amplitudes as one (or the maximum value) for all reflecting elements, i.e., $|{\theta}_{n}|=1, \forall n=1,\ldots,N$ 
\cite{Zheng2020DoubleIRS,zheng2020efficient,mei2021intelligent,Wu2019TWC}. 
Thus, given such unit-modulus constraint, we can represent any IRS reflection vector ${\bm \theta}$ as
\begin{align}\label{decomposition}
{\bm \theta}=e^{j\varphi}{\bar{\bm \theta}} \quad \Leftrightarrow \quad {\theta}_{n}=e^{j\varphi}{\bar{\theta}}_{n}, \qquad \forall n=1,\ldots,N
\end{align}
where ${\bar{\bm \theta}}\triangleq[{\bar{\theta}}_{1},{\bar{\theta}}_{2},\ldots,{\bar{\theta}}_{N}]^T$ represents the (passive) precoding vector with $|{\bar{\theta}}_{n}|=1, \forall n=1,\ldots,N$,
and $\varphi$ stands for the IRS's common phase-shift.
As will be demonstrated in Section~\ref{Passive Beamforming}, the IRS passive precoding gain is invariant to $\varphi$ and depends only on ${\bar{\bm \theta}}$.
As such, inspired by the decomposition structure in \eqref{decomposition}, we aim to exploit the IRS's common phase-shift $\varphi$ and precoding vector ${\bar{\bm \theta}}$ to simultaneously achieve transmit diversity for high-mobility users without CSI and passive precoding for low-mobility users with CSI, respectively.

\subsection{Communication Protocol}

\begin{figure}[!t]
	\centering
	\includegraphics[width=3.5in]{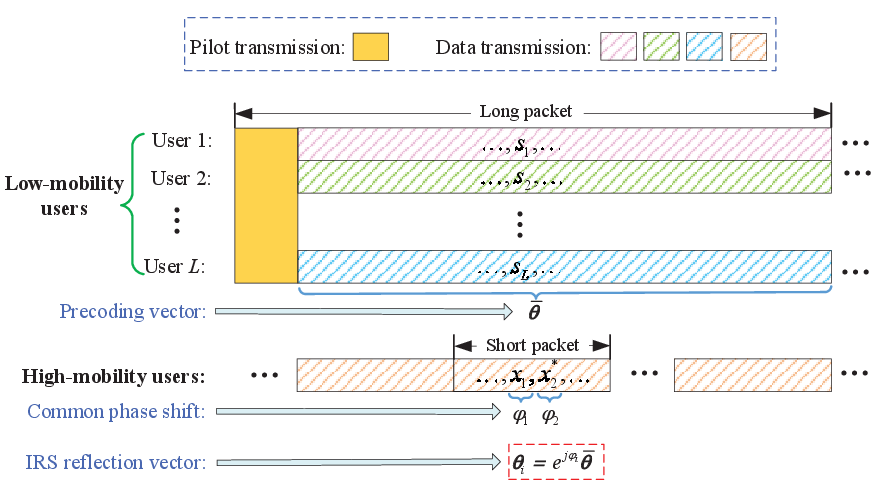}
	\setlength{\abovecaptionskip}{-3pt}
	\caption{Communication protocol for IRS-aided transmit diversity and active/passive precoding.}
	\label{protocol}
\end{figure}
Attentive to different channel conditions, we offer a practical communication protocol for simultaneously serving high-mobility and low-mobility users via short-packet MC and long-packet BC, respectively, as shown in Fig.~\ref{protocol}.
Specifically, for high-mobility users in ${\cal{K}}$ with a relatively small channel coherence time, the BS sends them common short packets.
In each packet, the common phase-shift $\varphi$ of the IRS is adjusted on a per-symbol basis in real time throughout the transmission to attain transmit diversity.
Furthermore, we can assume that the channels $\left\{{\bm g}_{k}^H, {\bm h}_{k}^H\right\}, \forall k\in{\cal{K}}$ remain constant within each short packet, e.g., by setting its duration less than the minimum channel coherence time of all high-mobility users.

On the other hand, for low-mobility users with longer channel coherence time, the BS sends independent long packets to each user in ${\cal{L}}$, where
each transmission packet comprises one pilot sequence followed by one long data frame, as depicted in Fig.~\ref{protocol}. 
We similarly assume that channels $\left\{{\bm g}_{l}^H, {\bm h}_{l}^H\right\}, \forall l\in{\cal{L}}$ remain constant within each long packet (e.g., by setting its duration less than the minimum channel coherence time of all low-mobility users).
Therefore, the precoding vector ${\bar{\bm \theta}}$ can remain constant throughout each long packet, ensuring a consistent passive precoding gain for low-mobility users.

\section{Proposed Scheme and Problem Formulation}\label{Model}
In this paper, we adopt linear transmit precoding at the BS, in which each data stream is allocated an independent precoding vector.
Accordingly, we can express the complex baseband transmitted signal at the BS as
\begin{align}\label{Tx_sig}
{\bm x}=\sum_{l=1}^{L} {\bm w}_l s_l+ {\bm w}_{L+1} {\tilde s}
\end{align}
where $s_l$ stands for the independent data symbol for low-mobility user $l$ with $l\in{\cal{L}}$, 
${\tilde s}$ denotes the common data symbol for high-mobility users in ${\cal{K}}$,
and ${\bm w}_l \in {\mathbb{C}^{ M\times 1}}$ and ${\bm w}_{L+1} \in {\mathbb{C}^{ M\times 1}}$ denote the corresponding transmit precoding vectors for symbols $s_l$ and ${\tilde s}$, respectively.
It is assumed that $\left\{s_l\right\}_{l=1}^L$ and ${\tilde s}$ are independent random
variables with zero mean and unit variance (i.e., normalized power).

\subsection{Transmit Diversity for High-mobility Users}\label{Transmit Diversity}
\rev{In this subsection, we investigate the transmit diversity design at the IRS-integrated BS without requiring any CSI of high-mobility users.}
Particularly, drawing inspiration from the classic Alamouti's scheme \cite{alamouti1998simple}, we develop a transmit diversity scheme aided by IRS\footnote{The proposed IRS-aided transmit diversity scheme in this paper can be extended to achieve a higher diversity order by adopting a more general space-time code design suitable for the configuration of a multi-antenna BS and multi-IRS (e.g., by partitioning the IRS into multiple small-size IRSs).}, using the communication protocol depicted in Fig.~\ref{protocol}
to deliver common information to high-mobility users in ${\cal{K}}$.
It's worth noting that, without the requirement for assuming channel reciprocity and CSI feedback, our proposed transmit diversity scheme at the IRS-integrated BS is applicable to time/frequency division duplexing (TDD/FDD). This versatility is particularly attractive for high-mobility communication applications in practice.

In the proposed IRS-aided transmit diversity scheme, the BS and IRS work collaboratively to  encode successive pairs of modulated data symbols under a new space-time code design with transmit precoding ${\bm w}_{L+1}$. As shown in Fig.~\ref{protocol}, we let ${\tilde s}_1$ and ${\tilde s}_2$ represent any pair of two successive modulated symbols, which are assumed to be independently taken from an $M$-ary (differential) phase-shift keying (PSK) constellation.\footnote{The (differential) PSK modulation employed in this paper can be further generalized to the (differential) amplitude/phase-shift keying (APSK) modulation which incorporates supplementary information into the (shared) signal amplitude of ${\tilde s}_1$ and ${\tilde s}_2$. Moreover, the CSI at the receiver side can also be dispensed with by using differential modulation.}
We design the space-time code for the joint transmit precoding of the BS and the common phase-shift of the IRS in Table.~\ref{coding}.
In specific, during the first symbol period, the BS sends ${\bm w}_{L+1}{\tilde s}_1$ (multiplexed with other data symbols $\left\{{\bm w}_ls_l^{(1)}\right\}_{l=1}^L$ for low-mobility users) and the IRS sets its reflection vector as
${\bm \theta}_1=e^{j\varphi_1}{\bar{\bm \theta}}$ and its common phase-shift as the phase difference between the two modulated symbols, i.e., $\varphi_1=\angle {\tilde s}_2-\angle {\tilde s}_1$.
During the second symbol period, the BS sends $-{\bm w}_{L+1}{\tilde s}_2^*$ (multiplexed with other data symbols $\left\{{\bm w}_l s_l^{(2)}\right\}_{l=1}^L$ for low-mobility users) and the IRS sets its reflection vector as 
${\bm \theta}_2=e^{j\varphi_2}{\bar{\bm \theta}}$ and its common phase-shift as $\varphi_2=\angle {\tilde s}_2-\angle {\tilde s}_1+\pi$. 
In Table~\ref{coding}, we further demonstrate the space-time code design for the proposed transmit diversity scheme at the IRS-integrated BS, comparing it with the transmit beamformed Alamouti's scheme \cite{Xiaoxiao2013physical} at the multi-antenna BS without IRS, where ${\bm w}_{L+1}$ and ${\bm w}'_{L+1}$ denote two orthogonal transmit precoding vectors.
\begin{table*}[]
	\centering
	\caption{The Space-time Code Comparison between IRS-aided Transmit Diversity and Transmit Beamformed Alamouti's Schemes.}\label{coding}
	\resizebox{0.9\textwidth}{!}{
		\begin{tabular}{c|cc||cc|}
			\cline{2-5}
			& \multicolumn{2}{c||}{IRS-aided transmit diversity scheme}                     & \multicolumn{2}{c|}{Transmit beamformed Alamouti's scheme \cite{Xiaoxiao2013physical}}           \\ 
			& \multicolumn{2}{c||}{(for the IRS-integrated BS)}                     & \multicolumn{2}{c|}{(for the multi-antenna BS without IRS)}           \\ \cline{1-5} 
			\multicolumn{1}{|c|}{Symbol index} & \multicolumn{1}{c|}{BS's transmit beam} & IRS's common phase-shift & \multicolumn{1}{c|}{BS's transmit beam 1} & BS's transmit beam 2 \\ \hline
			\multicolumn{1}{|c|}{1} & \multicolumn{1}{c|}{${\bm w}_{L+1}{\tilde s}_1$}    & $\varphi_1=\angle {\tilde s}_2-\angle {\tilde s}_1$~~~~~             & \multicolumn{1}{c|}{${\bm w}_{L+1}{\tilde s}_1$}    & ${\bm w}'_{L+1}{\tilde s}_2$    \\ \hline
			\multicolumn{1}{|c|}{2} & \multicolumn{1}{c|}{$-{\bm w}_{L+1}{\tilde s}_2^*$} & $\varphi_2=\angle {\tilde s}_2-\angle {\tilde s}_1+\pi$ & \multicolumn{1}{c|}{$-{\bm w}_{L+1}{\tilde s}_2^*$} & ${\bm w}'_{L+1}{\tilde s}_1^*$ \\ \hline
		\end{tabular}
	}
\end{table*}

Based on the proposed space-time code design, we can express the received signal at each high-mobility user over the first symbol period as:
\begin{align}\label{S1}
y_k^{(1)}=&\left( {\bm h}_{k}^H + {\bm g}_{k}^H {\rm diag}\left({\bm \theta}_1\right) {\bm R}  \right) \left(\sum_{l=1}^{L} {\bm w}_l s_l^{(1)}+ {\bm w}_{L+1} {\tilde s}_1\right)+n_k^{(1)}\notag\\
=&\left(  {\bm h}_{k}^H + {\bm g}_{k}^H {\rm diag}\left({\bm \theta}_1\right) {\bm R}  \right)  {\bm w}_{L+1} {\tilde s}_1\notag\\
&+
\underbrace{\left( {\bm h}_{k}^H + {\bm g}_{k}^H {\rm diag}\left({\bm \theta}_1\right) {\bm R} \right) \sum_{l=1}^{L} {\bm w}_l s_l^{(1)}+n_k^{(1)}}_{{\rm  interference-plus-noise:~}{\tilde n}_k^{(1)}}
\end{align}
where $n_k^{(1)}\sim {\mathcal N_c }(0, \sigma^2)$ is the zero-mean additive white Gaussian noise (AWGN) at high-mobility user $k$ and ${\tilde n}_k^{(1)}\sim {\mathcal N_c }(0, {\tilde \sigma}^2_k)$ is the equivalent interference-plus-noise with its variance given by 
\begin{align}\label{variance1}
{\tilde \sigma}^2_k&={\mathbb E}\left\{\sum_{l=1}^{L} \left|\left( {\bm h}_{k}^H + {\bm g}_{k}^H {\rm diag}\left({\bm \theta}_1\right) {\bm R} \right){\bm w}_l\right|^2\right\} +\sigma^2\notag\\
&=\sum_{l=1}^{L} {\bm w}_l^H {\mathbb E}\Big\{\left( {\bm h}_{k}^H + {\bm g}_{k}^H {\rm diag}\left({\bm \theta}_1\right) {\bm R} \right)^H\notag\\
 &\qquad\qquad\quad\left( {\bm h}_{k}^H + {\bm g}_{k}^H {\rm diag}\left({\bm \theta}_1\right) {\bm R} \right)\Big\}{\bm w}_l +\sigma^2\notag\\
&=\sum_{l=1}^{L} {\bm w}_l^H \left( \frac{\beta}{d_{k}^\alpha} {\bm I}_M + \frac{2\beta}{d_{k}^\alpha} {\bm R}^H{\bm R} \right)
{\bm w}_l +\sigma^2.
\end{align}
By substituting ${\bm \theta}_1=e^{j\varphi_1}{\bar{\bm \theta}}$ into \eqref{S1}, we have
\begin{align}\label{S1.1}
y_k^{(1)}&= \underbrace{{\bm h}_{k}^H   {\bm w}_{L+1} }_{{\tilde{h}_{k}}} {\tilde s}_1
+\underbrace{{\bm g}_{k}^H {\rm diag}\left({\bar{\bm \theta}}\right) {\bm R} {\bm w}_{L+1}}_{{\tilde{g}_{k}}} e^{j\varphi_1}{\tilde s}_1+ {\tilde n}_k^{(1)}\notag\\
&\stackrel{(a)}{=}{\tilde{h}_{k}} {\tilde s}_1 +{\tilde{g}_{k}} {\tilde s}_2+{\tilde n}_k^{(1)}
\end{align}
where ${\tilde{g}_{k}}$ and ${\tilde{h}_{k}}$ denote the effective gains of the IRS-reflected (i.e., BS$\rightarrow$IRS$\rightarrow$user~$k$) and BS$\rightarrow$user~$k$ channels, respectively, and $(a)$ holds providing that ${\tilde s}_2={\tilde s}_1 e^{j (\angle {\tilde s}_2-\angle {\tilde s}_1)}=e^{j\varphi_1} {\tilde s}_1$.

Following on, the received signal during the second symbol period at each high-mobility user is given by
\begin{align}\label{S2}
y_k^{(2)}=&\left( {\bm h}_{k}^H + {\bm g}_{k}^H {\rm diag}\left({\bm \theta}_2\right) {\bm R}  \right) \hspace{-0.05cm}\left(\hspace{-0.05cm}\sum_{l=1}^{L} {\bm w}_l s_l^{(2)}-{\bm w}_{L+1} {\tilde s}_2^*\hspace{-0.1cm}\right)
 \hspace{-0.1cm}+\hspace{-0.1cm}n_k^{(2)}\notag\\
 =&-\left(  {\bm h}_{k}^H + {\bm g}_{k}^H {\rm diag}\left({\bm \theta}_2\right) {\bm R}  \right)  {\bm w}_{L+1} {\tilde s}_2^*\notag\\
 &+
 \underbrace{\left( {\bm h}_{k}^H + {\bm g}_{k}^H {\rm diag}\left({\bm \theta}_2\right) {\bm R} \right) \sum_{l=1}^{L} {\bm w}_l s_l^{(2)}+n_k^{(2)}}_{{\rm  interference-plus-noise:~}{\tilde n}_k^{(2)}}
\end{align}
where $n_k^{(2)}\sim {\mathcal N_c }(0, \sigma^2)$ is the zero-mean AWGN at high-mobility user $k$ and ${\tilde n}_k^{(2)}\sim {\mathcal N_c }(0, {\tilde \sigma}^2_k)$ is the equivalent interference-plus-noise with its variance given in \eqref{variance1} by replacing ${\bm \theta}_1$ with ${\bm \theta}_2$. Similarly, by substituting ${\bm \theta}_2=e^{j\varphi_2}{\bar{\bm \theta}}$ into \eqref{S2}, we have
\begin{align}\label{S2.1}
y_k^{(2)}&= -\underbrace{{\bm h}_{k}^H   {\bm w}_{L+1} }_{{\tilde{h}_{k}}} {\tilde s}_2^*
-\underbrace{{\bm g}_{k}^H {\rm diag}\left({\bar{\bm \theta}}\right) {\bm R} {\bm w}_{L+1}}_{{\tilde{g}_{k}}} e^{j\varphi_2}{\tilde s}_2^*+ {\tilde n}_k^{(2)}\notag\\
&\stackrel{(b)}{=}-{\tilde{h}_{k}} {\tilde s}_2^* +{\tilde{g}_{k}} {\tilde s}_1^*+{\tilde n}_k^{(2)}
\end{align}
where $(b)$ holds providing that ${\tilde s}_1^*=-{\tilde s}_2^*e^{j (\angle {\tilde s}_2-\angle {\tilde s}_1+\pi)}=-e^{j\varphi_2}{\tilde s}_2^*$.

For the downlink MC, each high-mobility user focuses only on decoding ${\tilde s}_1$ and ${\tilde s}_2$ while treating other data symbols for low-mobility users as interference. 
As such, by letting ${\bm y}_k\triangleq\left[y_k^{(1)}, (y_k^{(2)})^*\right]^T$ represent the received signal vector associated with each transmitted pair of ${\tilde s}_1$ and ${\tilde s}_2$, \eqref{S1.1} and \eqref{S2.1} can be transformed into a more concise form as
\begin{align}\label{S1S2}
{\bm y}_k=\underbrace{\begin{bmatrix}
	{\tilde{h}_{k}}  &{\tilde{g}_{k}}\\
	{\tilde{g}_{k}}^*  &-{\tilde{h}_{k}}^*\end{bmatrix}}_{{\bm H}}
\underbrace{\begin{bmatrix}
	{\tilde s}_1 \\
	{\tilde s}_2 \end{bmatrix}}_{\tilde{\bm s}}+\underbrace{\begin{bmatrix}
	{\tilde n}_k^{(1)} \\
	\left({\tilde n}_k^{(2)}\right)^* \end{bmatrix}}_{{\tilde{\bm n}}_k} 
\end{align}
where ${\bm H}$ stands for the equivalent channel matrix, ${\tilde{\bm s}}$ denotes the modulated symbol vector, and ${\tilde{\bm n}}_k$ denotes the interference-plus-noise vector.
Then, left-multiplying the received signal vector ${\bm y}_k$ in \eqref{S1S2} with ${\bm H}^H$, we obtain
\begin{align}\label{decoupled}
{\bar{\bm y}_k}={\bm H}^H{\bm y}_k={\bm H}^H {\bm H}{\tilde{\bm s}}+{\bar{\bm n}_k}
\end{align}
with ${\bar{\bm y}}\triangleq\left[{\bar y}_1,{\bar y}_2\right]^T$ and ${\bar{\bm n}}_k\triangleq{\bm H}^H {\tilde{\bm n}}_k$, where we have ${\bm H}^H {\bm H}=\left(|{\tilde{h}_{k}}|^2+|{\tilde{g}_{k}}|^2\right){\bm I}_2$
and it can be verified that ${\bar{\bm n}}_k$ is the equivalent interference-plus-noise vector with ${\bar{\bm n}}\sim {\mathcal N_c }({\bm 0}, (|{\tilde{h}_{k}}|^2+|{\tilde{g}_{k}}|^2){\tilde \sigma}^2_k{\bm I}_2)$.
Consequently, the received SINR at each high-mobility user is given by
\begin{align}\label{SNR}
&\text{SINR}_k=\frac{|{\tilde{h}_{k}}|^2+|{\tilde{g}_{k}}|^2}{{\tilde \sigma}^2_k}\notag\\
&=\frac{|{\bm h}_{k}^H {\bm w}_{L+1}|^2+|{\bm g}_{k}^H {\rm diag}\left({\bar{\bm \theta}}\right) {\bm R} {\bm w}_{L+1}|^2}{\sum\limits_{l=1}^{L} {\bm w}_l^H \left( \frac{\beta}{d_{k}^\alpha} {\bm I}_M + \frac{2\beta}{d_{k}^\alpha} {\bm R}^H{\bm R} \right)
	{\bm w}_l +\sigma^2},\forall k\in{\cal{K}}.
\end{align}

\subsection{Active/Passive Precoding for Low-mobility Users}\label{Passive Beamforming}
\rev{While serving high-mobility users through the IRS-aided transmit diversity, the IRS-integrated BS also concurrently performs conventional active/passive precoding to serve low-mobility users with their CSI known.}\footnote{\rev{For the active/passive precoding design at the IRS-integrated BS, the required CSI for low-mobility users can be obtained using existing cascaded channel estimation schemes (see, e.g., \cite{zheng2019intelligent,zheng2020intelligent,yang2019intelligent,Zheng2020Fast,You2020Fast,wei2021channel,he2019cascaded}).}}
Specifically, with a given common phase-shift which helps achieve transmit diversity for high-mobility users, the precoding vector ${\bar{\bm \theta}}$ in \eqref{decomposition} can be utilized for low-mobility users with appropriate design, as shown in Fig.~\ref{system}.
Recall that the common phase-shift $\varphi$ is pre-established for achieving transmit diversity and thus is known and can be compensated at the IRS-integrated BS in advance.

As such, the signal received at low-mobility user $l$ from both the BS$\rightarrow$user and BS$\rightarrow$IRS$\rightarrow$user
channels can be represented as
\begin{align}\label{Low_model}
y_l=&\left( {\bm h}_{l}^H+ {\bm g}_{l}^H  {\rm diag}\left({\bm \theta}\right) {\bm R}   \right) \left(\sum_{j=1}^{L} {\bm w}_j s_j+ {\bm w}_{L+1} {\tilde s}\right)+n_l\notag\\
=&\hspace{-0.1cm}\left( {\bm h}_{l}^H\hspace{-0.1cm}+\hspace{-0.1cm} e^{j\varphi}{\bm g}_{l}^H  {\rm diag}\left({\bar{\bm \theta}}\right) {\bm R}   \right){\bm w}_l s_l \hspace{-0.1cm}+\hspace{-0.1cm}\left( {\bm h}_{l}^H\hspace{-0.1cm}+\hspace{-0.1cm} e^{j\varphi}{\bm g}_{l}^H  {\rm diag}\left({\bar{\bm \theta}}\right) {\bm R}   \right)\notag\\
&\left(\sum_{j\ne l}^L {\bm w}_j s_j+ {\bm w}_{L+1} {\tilde s}\right)+n_l
\hspace{-0.1cm}\end{align}
where $n_l\sim {\mathcal N_c }(0, \sigma^2)$ stands for the AWGN at low-mobility user $l$ with $\sigma^2$ representing the equivalent noise power.
Furthermore, we let ${\bar{\bm g}}_{l}^H \left({\bar{\bm \theta}}\right) \triangleq e^{j\varphi}{\bm g}_{l}^H  {\rm diag}\left({\bar{\bm \theta}}\right) {\bm R} {\bm w}_l, \forall l\in{\cal{L}}$, which is the equivalent complex-valued gain of the IRS-reflected channel for low-mobility user $l$. 
In particular, for any previously assumed precoding vector ${\bar{\bm \theta}}$, we can obtain
\begin{align}
\left|{\bar{\bm g}}_{l}^H \left({\bar{\bm \theta}}\right)\right|^2&=\left|e^{j\varphi}{\bm g}_{l}^H  {\rm diag}\left({\bar{\bm \theta}}\right) {\bm R} {\bm w}_l\right|^2\notag\\
&=\left|{\bm g}_{l}^H  {\rm diag}\left({\bar{\bm \theta}}\right) {\bm R} {\bm w}_l\right|^2.
\end{align}
This indicates that the IRS passive precoding gain in any (channel) direction to any low-mobility user remains unchanged when dynamically altering the common phase-shift $\varphi$.
According to \eqref{Low_model}, the SINR of low-mobility user $l$ is given by 
\begin{align}
\text{SINR}_l\hspace{-0.1cm}=\hspace{-0.1cm}\frac{\left|\left( {\bm h}_{l}^H+ e^{j\varphi}{\bm g}_{l}^H  {\rm diag}\left({\bar{\bm \theta}}\right) {\bm R}  \right) {\bm w}_l\right|^2}
{\sum\limits_{j\ne l}^{L+1}\left|\left( {\bm h}_{l}^H+ e^{j\varphi}{\bm g}_{l}^H  {\rm diag}\left({\bar{\bm \theta}}\right) {\bm R}  \right) {\bm w}_j\right|^2\hspace{-0.1cm}+\hspace{-0.1cm}\sigma^2}.
\end{align}

\subsection{Problem Formulation}
In order to minimize the total transmit power at the BS, we jointly optimize the reflect precoding at the IRS along with transmit precoding at the BS under individual SINR constraints at all users.
For simplicity of notation, we let $ {\bm W} =\left[{\bm w}_1,\cdots,{\bm w}_L, {\bm w}_{L+1}\right]$ denote the transmit precoding matrix at the BS.
As such, the problem is formulated as
\begin{align}
\hspace{-0.4cm}\text{(P1):}
& \underset{ {\bm W},{\bar{\bm \theta}}}{\text{min}}
& &\hspace{-0.35cm}  \sum_{j=1}^{L+1} \left\|{\bm w}_j\right\|^2 \label{obj_P1} \\
& \text{s.t.} & &\hspace{-0.35cm}  \frac{\left|\left( {\bm h}_{l}^H+ e^{j\varphi}{\bm g}_{l}^H  {\rm diag}\left({\bar{\bm \theta}}\right) {\bm R}  \right) {\bm w}_l\right|^2}
{\sum\limits_{j\ne l}^{L+1}\hspace{-0.1cm}\left|\hspace{-0.08cm}\left( {\bm h}_{l}^H\hspace{-0.1cm}+\hspace{-0.1cm} e^{j\varphi}{\bm g}_{l}^H  {\rm diag}\hspace{-0.1cm}\left(\hspace{-0.03cm}{\bar{\bm \theta}}\hspace{-0.03cm}\right)\hspace{-0.05cm} {\bm R} \hspace{-0.05cm} \right)\hspace{-0.1cm} {\bm w}_j\right|^2\hspace{-0.1cm}+\hspace{-0.1cm}\sigma^2}\ge \gamma_l, \forall l\hspace{-0.1cm}\in\hspace{-0.1cm}{\cal{L}} \hspace{-0.2cm}\label{con1_P1}\\
& & & \hspace{-0.35cm}\frac{|{\bm h}_{k}^H {\bm w}_{L+1}|^2\hspace{-0.1cm}+\hspace{-0.1cm}|{\bm g}_{k}^H {\rm diag}\hspace{-0.1cm}\left(\hspace{-0.03cm}{\bar{\bm \theta}}\hspace{-0.03cm}\right)\hspace{-0.1cm} {\bm R} {\bm w}_{L+1}|^2\hspace{-0.1cm}}{\sum\limits_{l=1}^{L}\hspace{-0.1cm} {\bm w}_l^H\hspace{-0.1cm} \left(\hspace{-0.05cm} \frac{\beta}{d_{k}^\alpha} {\bm I}_M \hspace{-0.1cm}+\hspace{-0.1cm} \frac{2\beta}{d_{k}^\alpha} {\bm R}^H\hspace{-0.05cm}{\bm R}\hspace{-0.05cm} \right)\hspace{-0.1cm}
	{\bm w}_l \hspace{-0.1cm}+\hspace{-0.1cm}\sigma^2}\hspace{-0.1cm}\ge \hspace{-0.1cm}{\tilde \gamma}_{L+1},\hspace{-0.1cm} \forall k\hspace{-0.1cm}\in{\cal{K}}\hspace{-0.1cm}\label{con2_P1}\\
& & &\hspace{-0.35cm}|{\theta}_{n}|=1, \forall n=1,\ldots,N \label{con3_P1}
\end{align}
where $\gamma_l>0, l \in {\cal{L}}$ stands for the minimum individual SINR requirement of each low-mobility user and ${\tilde \gamma}_{L+1}>0$ stands for the minimum SINR requirement of the common information to all high-mobility users in ${\cal{K}}$. Recall that for high-mobility users, no CSI is assumed at the IRS-integrated BS, i.e., the instantaneous CSI of ${\bm h}_{k}$ and ${\bm g}_{k}$ is unavailable under the constraint of \eqref{con2_P1}. Alternatively, based on the channel distributions given in \eqref{Rayleigh}, we turn to take the expectation of SINR for each high-mobility user in \eqref{SNR} over ${\bm h}_{k}$ and ${\bm g}_{k}$, which is given by 
\begin{align}\label{SNR2}
&\hspace{-0.3cm}\overline{\text{SINR}}_k=\frac{{\mathbb E}\left\{|{\bm h}_{k}^H {\bm w}_{L+1}|^2+|{\bm g}_{k}^H {\rm diag}\left({\bar{\bm \theta}}\right) {\bm R} {\bm w}_{L+1}|^2\right\}}{\sum\limits_{l=1}^{L} {\bm w}_l^H \left( \frac{\beta}{d_{k}^\alpha} {\bm I}_M + \frac{2\beta}{d_{k}^\alpha} {\bm R}^H{\bm R} \right)
	{\bm w}_l +\sigma^2}\notag\\
&\hspace{-0.3cm}=\frac{{\bm w}_{L+1}^H \left(  {\bm I}_M + 2 {\bm R}^H{\bm R} \right)
	{\bm w}_{L+1}}{\sum\limits_{l=1}^{L} {\bm w}_l^H \left( {\bm I}_M + 2 {\bm R}^H{\bm R} \right)
	{\bm w}_l +{d_{k}^\alpha}\sigma^2/{\beta}}
,\quad \forall k\in{\cal{K}}.
\end{align}
Accordingly, let $\gamma_{L+1}>0$ denote the minimum average SINR requirement of all high-mobility users in ${\cal{K}}$, problem (P1) is thus reformulated as
\begin{align}
\hspace{-0.4cm}\text{(P2):}
& \underset{ {\bm W},{\bar{\bm \theta}}}{\text{min}}
& &\hspace{-0.35cm}  \sum\limits_{j=1}^{L+1} \left\|{\bm w}_j\right\|^2 \label{obj_P2} \\
& \text{s.t.} & &\hspace{-0.35cm} \frac{\left|\left( {\bm h}_{l}^H+ e^{j\varphi}{\bm g}_{l}^H  {\rm diag}\left({\bar{\bm \theta}}\right) {\bm R}  \right) {\bm w}_l\right|^2}
{\sum\limits_{ j\ne l}^{L+1} \hspace{-0.1cm} \left| \hspace{-0.05cm} \left( {\bm h}_{l}^H \hspace{-0.13cm} + \hspace{-0.1cm} e^{j\varphi}{\bm g}_{l}^H  {\rm diag} \hspace{-0.1cm} \left({\bar{\bm \theta}}\right) \hspace{-0.1cm} {\bm R}  \right) \hspace{-0.1cm} {\bm w}_j\right|^2 \hspace{-0.1cm} + \hspace{-0.1cm} \sigma^2} \hspace{-0.1cm} \ge \hspace{-0.1cm} \gamma_l, \forall l\in{\cal{L}} \label{con1_P2}\\ 
& & &\hspace{-0.35cm} \frac{{\bm w}_{L+1}^H \left(  {\bm I}_M + 2 {\bm R}^H{\bm R} \right)
	{\bm w}_{L+1}}{\sum\limits_{l=1}^{L} \hspace{-0.05cm} {\bm w}_l^H \hspace{-0.1cm} \left( {\bm I}_M \hspace{-0.1cm} + \hspace{-0.1cm} 2 {\bm R}^H{\bm R} \right)
	{\bm w}_l \hspace{-0.1cm} + \hspace{-0.1cm} {d_{k}^\alpha}\sigma^2 \hspace{-0.05cm} / \hspace{-0.05cm} {\beta}} \hspace{-0.1cm} \ge \hspace{-0.1cm} \gamma_{L+1}, \forall k \hspace{-0.05cm} \in \hspace{-0.05cm} {\cal{K}}\label{con2_P2}\\
& & &\hspace{-0.35cm}|{\bar{\theta}}_{n}|=1, \forall n=1,\ldots,N. \label{con3_P2}
\end{align}
By taking a closer look, we find that the constraints in \eqref{con2_P2} can be reduced to
\begin{align}\label{con4_P2}
\frac{{\bm w}_{L+1}^H \left(  {\bm I}_M + 2 {\bm R}^H{\bm R} \right)
	{\bm w}_{L+1}}{\sum\limits_{l=1}^{L} {\bm w}_l^H \left( {\bm I}_M + 2 {\bm R}^H{\bm R} \right)
	{\bm w}_l +{\bar \sigma}^2_{L+1}}\ge \gamma_{L+1}
\end{align}
where ${\bar \sigma}^2_{L+1}\triangleq\underset{k\in{\cal{K}}}{\text{max}} ~ \frac{{d_{k}^\alpha}\sigma^2}{\beta} $.
Although the objective function in \eqref{obj_P2} is convex, it is tricky to solve (P2) because of the non-convex constraints. Specifically, the unit-modulus constraint in \eqref{con3_P2} is non-convex, and 
 the transmit precoding ${\bm W}$ and reflect precoding ${\bar{\bm \theta}}$ are coupled, which thus need to be jointly optimized. 
In addition, it can be observed from \eqref{con1_P2} and \eqref{con4_P2} that all the low-mobility and high-mobility users are coupled by their mutual interference, which makes the problem even more challenging to solve.
While there is no established approach for optimally addressing such a non-convex optimization problem like (P2), we employ the AO technique, as elaborated in the next section, to handle it efficiently.

\section{Proposed Solution to Problem (P2)}\label{Solution}
In this section, we offer an AO algorithm to solve (P2).
Specifically, we alternately optimize the transmit precoding ${\bm W}$ at the BS and the reflect precoding ${\bar{\bm \theta}}$ at the IRS iteratively, until convergence is achieved.
For any given reflect precoding ${\bar{\bm \theta}}$, the superimposed channel from the BS to each low-mobility user can be expressed as
\begin{align}\label{combined_channel}
{\bar {\bm h}}_{l}^H= {\bm h}_{l}^H+ e^{j\varphi}{\bm g}_{l}^H  {\rm diag}\left({\bar{\bm \theta}}\right) {\bm R}
\end{align}
and thus problem (P2) reduces to
\begin{align}
\text{(P3):}
& \underset{ {\bm W}}{\text{min}}
& &  \sum\limits_{j=1}^{L+1} \left\|{\bm w}_j\right\|^2 \label{obj_P3} \\
& \text{s.t.} & &  \frac{\left|{\bar {\bm h}}_{l}^H {\bm w}_l\right|^2}
{\sum\limits_{j\ne l}^{L+1}\left|{\bar {\bm h}}_{l}^H  {\bm w}_j\right|^2+\sigma^2}\ge \gamma_l, \forall l\in{\cal{L}} \label{con1_P3}\\
& & & \frac{{\bm w}_{L+1}^H \left(  {\bm I}_M + 2 {\bm R}^H{\bm R} \right)
	{\bm w}_{L+1}}{\sum\limits_{l=1}^{L} {\bm w}_l^H \left( {\bm I}_M + 2 {\bm R}^H{\bm R} \right)
	{\bm w}_l +{\bar \sigma}^2_{L+1}} \hspace{-0.1cm} \ge \hspace{-0.1cm} \gamma_{L+1}.\label{con2_P3}
\end{align}
Although problem (P3) shares a similar form to the conventional power minimization problem in the multiuser multiple-input single-output (MISO) downlink broadcast system, it has an additional average SINR constraint for high-mobility users as shown in  \eqref{con2_P3}. It is found that after reformulating problem (P3) as a convex problem, we can apply the semidefinite program (SDP) \cite{bengtsson2001book} or second-order cone program (SOCP) \cite{Wiesel2006Linear}  to solve it efficiently.
Nevertheless, this numerical solution has few insights into the structure of its optimal/suboptimal solution. To tackle this problem, we present the optimal MMSE and suboptimal ZF transmit precoding in the following.
\subsection{Transmit Precoding Optimization}\label{Transmit Beamforming}
\subsubsection{MMSE Transmit Precoding}
To fully exploit the transmit precoding at the BS, we first apply the optimal MMSE criterion to cope with
interference among low-mobility and high-mobility users. 
For simplicity of notation, we let
${\bm Q}_l \triangleq  \frac{1}{\sigma^2}{\bar {\bm h}}_{l} {\bar {\bm h}}_{l}^H, \forall l\in{\cal{L}}$ and ${\bm Q}_{L+1} \triangleq  \frac{1}{{\bar \sigma}^2_{L+1}}\left({\bm I}_M + 2 {\bm R}^H{\bm R}\right)$ denote the channel correlation matrices for low-mobility and high-mobility users, respectively. By unifying the constraints in \eqref{con1_P3} and \eqref{con2_P3},
problem (P3) can be equivalently expressed as 
\begin{align}
\text{(P3.1):}
& \underset{ {\bm W}}{\text{min}}
& & \hspace{-0.05cm}  \sum_{j=1}^{L+1} \left\|{\bm w}_j\right\|^2 \label{obj_P3.1} \\
& \hspace{-0.5cm} \text{s.t.} & & \hspace{-1cm} \frac{1}{\gamma_j} {\bm w}_j^H  {\bm Q}_j 	{\bm w}_j 
-  \sum_{l\ne j}^{L+1}{\bm w}_l^H  {\bm Q}_j  {\bm w}_l
\ge 1 , \forall j\in{\cal{J}} \label{con1_P3.1} 
\end{align}
where ${\cal{J}}\triangleq{\cal{L}}\cup \{L+1\}$. 
It is noted that the inequality constraints specified in \eqref{con1_P3.1} meet with equality at the optimum;
otherwise, one could scale down the optimal solution ${\bm W}$ to ensure the equality, which would lead to a decrease in the objective function and contradict the optimality.
In particular, for a convex optimization problem, we may explore its dual to reveal the structure of its optimal solution.
We will demonstrate this point by revealing that the Lagrangian dual of problem (P3.1) holds an engineering interpretation, which is commonly referred to as uplink-downlink duality.
Accordingly, the Lagrange multiplier function is defined as
\begin{align}
&{\cal L} \left({\bm W} , {\bm \lambda}\right) \triangleq \sum_{j=1}^{L+1} \left\|{\bm w}_j  \right\|^2 \nonumber \\
& \hspace{1.5cm} -\sum_{j=1}^{L+1} \hspace{-0.1cm} \lambda_j  \left( \frac{1}{\gamma_j} {\bm w}_j^H  {\bm Q}_j 	{\bm w}_j 
 \sum_{l\ne j}^{L+1} \hspace{-0.1cm} {\bm w}_l^H  {\bm Q}_j  {\bm w}_l 
-  1 \hspace{-0.1cm} \right)\label{Lagrange1}\\
&\stackrel{(c)}{=}\sum_{j=1}^{L+1} \lambda_j+ \sum_{j=1}^{L+1} {\bm w}_j^H \left({\bm I}_M + \sum_{l\ne j}^{L+1} \lambda_l {\bm Q}_l - \frac{\lambda_j}{\gamma_j} {\bm Q}_j
\right)  {\bm w}_j \label{Lagrange2}
\end{align}
where $\lambda_j \ge 0$ denotes the Lagrange multiplier corresponding to the $j$-th SINR constraint with $j\in{\cal{J}}$ and $(c)$ is obtained via rearranging the terms of \eqref{Lagrange2}.
Therefore, the dual objective is given by 
\begin{align}
g\left({\bm \lambda}\right) = \underset{ {\bm W}}{\text{min}} ~{\cal L}\left({\bm W}, {\bm \lambda}\right).
\end{align}
If ${\bm I}_M + \sum_{l\ne j}^{L+1} \lambda_l {\bm Q}_l - \frac{\lambda_j}{\gamma_j} {\bm Q}_j$ is not a positive semi-definite matrix, then it is easy to find that there exists a set of ${\bm w}_j^H$ that would make $g\left({\bm \lambda}\right)=-\infty$. 
Thus, the Lagrangian dual of problem (P3.1), which is the maximum of $g\left({\bm \lambda}\right)$, is
\begin{align}
\text{(P3.2):}
& \underset{ {\bm \lambda}}{\text{max}}
& &  \sum_{j=1}^{L+1} \lambda_j \label{obj_P3.2} \\
& \hspace{-0.5cm} \text{s.t.} & & \hspace{-0.5cm} {\bm I}_M + \sum_{l=1}^{L+1} \lambda_l {\bm Q}_l  \succeq \left(1+\frac{1}{\gamma_j}\right) \lambda_j{\bm Q}_j, \forall j\in{\cal{J}}. \label{con1_P3.2} 
\end{align}
It is proven that with strong duality \cite{Wiesel2006Linear}, the optimal value of the original downlink problem (P3.1) is identical to that of the dual problem (P3.2), which can be further written as
\begin{align}
\text{(P3.3):}
& \underset{ {\bm \lambda}, {\hat{\bm W}}}{\text{min}}
& &  \sum_{j=1}^{L+1} \lambda_j \label{obj_P3.3} \\
& \text{s.t.} & & \frac{ \lambda_j {\hat{\bm w}}_j^H {\bm Q}_j {\hat{\bm w}}_j  }{ \sum_{l\ne j}^{L+1} \lambda_l {\hat{\bm w}}_j^H {\bm Q}_l {\hat{\bm w}}_j  + {\hat{\bm w}}_j^H {\hat{\bm w}}_j } \ge \gamma_j
\label{con1_P3.3} 
\end{align}
where we identify $\lambda_j$ as the dual uplink power (with normalized dual uplink noise covariance) and $ {\hat{\bm W}} =\left[{\hat{\bm w}}_1,\cdots,{\hat{\bm w}}_L, {\hat{\bm w}}_{L+1}\right]$ is the receive beamforming matrix at the BS.
Clearly, problem (P3.3) is a virtual normalized uplink problem, which can be optimally solved with the receive beamforming design based on the MMSE. Moreover, it can be verified that ${\bm w}_ j$ and ${\hat{\bm w}}_j$ are scaled versions of each other.

Next, we take the differentiation of  ${\cal L}\left({\bm W}, {\bm \lambda}\right) $ in \eqref{Lagrange1} with respect to ${\bm w}_j$, yielding
\begin{align}
\frac{\partial{\cal L}\left({\bm W}, {\bm \lambda}\right)}{\partial{\bm w}_j}&={\bm w}_j+
\sum_{l\ne j}^{L+1} \lambda_l {\bm Q}_l  {\bm w}_j
- \frac{\lambda_j}{\gamma_j}  {\bm Q}_j 	{\bm w}_j , j\in{\cal{J}}.
\end{align}
By letting $\frac{\partial{\cal L}\left({\bm W}, {\bm \lambda}\right)}{\partial{\bm w}_j}=0$, we have 
\begin{align}
&\quad {\bm w}_j+
\sum_{l=1}^{L+1} \lambda_l {\bm Q}_l  {\bm w}_j
=  \lambda_j {\bm Q}_j  {\bm w}_j+
\frac{\lambda_j}{\gamma_j}  {\bm Q}_j 	{\bm w}_j\label{diff1}
\end{align}
where the term $\lambda_j {\bm Q}_j  {\bm w}_j$ is added to both sides in \eqref{diff1}. For $j\in{\cal{L}}$, by substituting ${\bm Q}_j \triangleq  \frac{1}{\sigma^2} {\bar {\bm h}}_{j} {\bar {\bm h}}_{j}^H$ into \eqref{diff1}, we have
\begin{align}
&\left( \hspace{-0.1cm} {\bm I}_M \hspace{-0.1cm} + \hspace{-0.1cm}
\sum_{l=1}^{L} \hspace{-0.1cm} \frac{\lambda_l}{\sigma^2} {\bar {\bm h}}_{l} {\bar {\bm h}}_{l}^H \hspace{-0.1cm} + \hspace{-0.1cm} \lambda_{L+1}{\bm Q}_{L+1} \hspace{-0.1cm} \right) \hspace{-0.1cm} {\bm w}_j \hspace{-0.05cm}
= \hspace{-0.05cm} \frac{\lambda_j}{\sigma^2} \left(1 \hspace{-0.1cm} + \hspace{-0.1cm}
\frac{1}{\gamma_j} \right) {\bar {\bm h}}_{j} {\bar {\bm h}}_{j}^H 	{\bm w}_j\notag\\
&\Rightarrow
{\bm w}_j=\left({\bm I}_M+
\sum_{l=1}^{L} \frac{\lambda_l}{\sigma^2} {\bar {\bm h}}_{l} {\bar {\bm h}}_{l}^H +\lambda_{L+1}{\bm Q}_{L+1} \right)^{-1}
{\bar {\bm h}}_{j} \notag\\
&\qquad\qquad\times
\underbrace{\frac{\lambda_j}{\sigma^2}\left(1+
\frac{1}{\gamma_j} \right) {\bar {\bm h}}_{j}^H 	{\bm w}_j}_{=\text{scalar}}, \quad j\in{\cal{L}}.
\end{align}
In other words, the optimal precoding vector ${\bm w}_j$ for $j\in{\cal{L}}$ can be given by
\begin{align}\label{beamforming_j}
{\bm w}_j^\star \hspace{-0.05cm} = \hspace{-0.05cm} \sqrt{p_j}\underbrace{\frac{\left({\bm I}_M \hspace{-0.05cm} + \hspace{-0.05cm}
	\sum_{l=1}^{L}\frac{\lambda_l}{\sigma^2} {\bar {\bm h}}_{l} {\bar {\bm h}}_{l}^H +\lambda_{L+1}{\bm Q}_{L+1} \right)^{-1}
	{\bar {\bm h}}_{j}}{\left\| \left({\bm I}_M \hspace{-0.1cm} + \hspace{-0.1cm}
	\sum_{l=1}^{L} \frac{\lambda_l}{\sigma^2} {\bar {\bm h}}_{l} {\bar {\bm h}}_{l}^H \hspace{-0.1cm} + \hspace{-0.1cm} \lambda_{L+1}{\bm Q}_{L+1} \right)^{-1} \hspace{-0.1cm}
	{\bar {\bm h}}_{j}\right\|}}_{\triangleq {\bar{\bm w}}_j: \text{normalized precoding direction}}
\end{align}
where $\sqrt{p_j}$ denotes the transmit precoding power and ${\bar{\bm w}}_j$ denotes the normalized
transmit precoding direction for low-mobility user $j\in{\cal{L}}$.

\newcounter{MYtempeqncnt}
\setcounter{MYtempeqncnt}{\value{equation}} %
\setcounter{equation}{47} %
\begin{figure*}[!b]
	\vspace*{4pt}
	\hrulefill
	\begin{align}\label{fixpeq}
		\hspace{-0.35cm}\left\{
		\begin{aligned}
			&\lambda_j= \frac{1}{\left(1+ \frac{1}{\gamma_j} \right){\bar {\bm h}}_{j}^H\left({\bm I}_M+
				\sum_{l=1}^{L} \frac{\lambda_l}{\sigma^2} {\bar {\bm h}}_{l} {\bar {\bm h}}_{l}^H +\lambda_{L+1}{\bm Q}_{L+1} \right)^{-1}
				{\bar {\bm h}}_{j} }, \quad j\in{\cal{L}}\\
			&\lambda_{L+1}=\frac{1}{\left(1+ \frac{1}{\gamma_{L+1}} \right) {\bar \lambda}_{\text{max}} \left[\left({\bm I}_M+
				\sum_{l=1}^{L} \frac{\lambda_l}{\sigma^2} {\bar {\bm h}}_{l} {\bar {\bm h}}_{l}^H +\lambda_{L+1}{\bm Q}_{L+1} \right)^{-1}
				{\bm Q}_{L+1}\right] }, \quad j=L+1\hspace{-0.2cm}
		\end{aligned}
		\right..
	\end{align}
\end{figure*}
\setcounter{equation}{\value{MYtempeqncnt}}

On the other hand, for $j=L+1$, we have
\begin{align}\label{lambdaL}
&\frac{1}{\lambda_{L+1}}{\bm w}_{L+1}=\left(1+ \frac{1}{\gamma_{L+1}} \right)\notag\\
\hspace{-0.1cm}\times&\left({\bm I}_M+
\sum_{l=1}^{L} \frac{\lambda_l}{\sigma^2} {\bar {\bm h}}_{l} {\bar {\bm h}}_{l}^H +\lambda_{L+1}{\bm Q}_{L+1} \right)^{-1}
  {\bm Q}_{L+1} 	{\bm w}_{L+1}.\hspace{-0.1cm}
\end{align}
Equation \eqref{lambdaL} implies that ${\bm w}_{L+1}$ should be selected as one of the eigenvectors of the matrix $\left(1+ \frac{1}{\gamma_{L+1}} \right) \left({\bm I}_M+
\sum_{l=1}^{L} \frac{\lambda_l}{\sigma^2} {\bar {\bm h}}_{l} {\bar {\bm h}}_{l}^H +\lambda_{L+1}{\bm Q}_{L+1} \right)^{-1} {\bm Q}_{L+1}$ and $\frac{1}{\lambda_{L+1}}$ is the associated eigenvalue.
As such, the optimal precoding vector ${\bm w}_{L+1}$ can be given by
\begin{align}\label{beamforming_L}
{\bm w}_{L+1}^\star=\sqrt{p_{L+1}}{\bar{\bm w}}_{L+1}
\end{align}
where 
$
{\bar{\bm w}}_{L+1} \hspace{-0.1cm} = \hspace{-0.1cm} {\mathbb P}\hspace{-0.07cm}\left\{\hspace{-0.15cm} \left(\hspace{-0.1cm}{\bm I}_M\hspace{-0.1cm}+\hspace{-0.1cm}
\sum\limits_{l=1}^{L} \frac{\lambda_l}{\sigma^2} {\bar {\bm h}}_{l} {\bar {\bm h}}_{l}^H \hspace{-0.07cm}+\hspace{-0.07cm}\lambda_{L+1}{\bm Q}_{L+1} \hspace{-0.07cm}\right)^{-1}
\hspace{-0.15cm}{\bm Q}_{L+1} \hspace{-0.1cm}\right\} 
$
is the normalized principal eigenvector (i.e., normalized transmit precoding direction) and $p_{L+1}$ denotes the transmit precoding power for high-mobility users.
Note that there are $L+1$ unknown precoding powers in \eqref{beamforming_j} and \eqref{beamforming_L}, which can be computed by leveraging the constraints in \eqref{con1_P3.1} that hold with equality at the optimal solution, i.e., 
\begin{align}\label{eq_opt}
&\quad \frac{1}{\gamma_j} {\bm w}_j^H  {\bm Q}_j 	{\bm w}_j 
-  \sum_{l\ne j}^{L+1}{\bm w}_l^H  {\bm Q}_j  {\bm w}_l
= 1 , \quad \forall j\in{\cal{J}}.
\end{align}
By substituting \eqref{beamforming_j} and \eqref{beamforming_L} into \eqref{eq_opt}, we have
\begin{align}
\hspace{-0.3cm}\left\{
\begin{aligned}
&\hspace{-0.15cm}\frac{p_j}{\sigma^2\gamma_j } \left|{\bar {\bm h}}_{j}^H	{\bar{\bm w}}_j \right|^2
-  \sum_{l\ne j}^{L+1} \frac{p_l}{\sigma^2} \left|{\bar {\bm h}}_{j}^H	{\bar{\bm w}}_l \right|^2
= 1 , j\in{\cal{L}}\\
&\hspace{-0.15cm}\frac{p_{L+1}}{\gamma_{L+1} } {\bar{\bm w}}_{L+1}^H  {\bm Q}_{L+1}  {\bar{\bm w}}_{L+1}
\hspace{-0.1cm}-\hspace{-0.1cm}  \sum_{l=1}^{L} p_l{\bar{\bm w}}_l^H  {\bm Q}_{L+1}  {\bar{\bm w}}_l
\hspace{-0.1cm}=\hspace{-0.1cm} 1 , j\hspace{-0.1cm}=\hspace{-0.1cm}L\hspace{-0.1cm}+\hspace{-0.1cm}1\hspace{-0.2cm}
\end{aligned}
\right..\hspace{-0.2cm}
\end{align}
As such, given the transmit precoding directions in \eqref{beamforming_j} and \eqref{beamforming_L}, we have $L+1$ linear equations for $L+1$ precoding powers, which can be written in a concise form as ${\bar{\bm Q}} {\bm p}={\bm 1}_{L+1}$,
where ${\bm p}=\left[p_1, \ldots, p_L, p_{L+1}\right]^T$ is the transmit precoding power vector and ${\bar{\bm Q}}$ is an $(L+1)\times (L+1)$ matrix with its 
$(\imath, \jmath )$-th element given by
\begin{align}
\left[{\bar{\bm Q}}\right]_{\imath, \jmath}\hspace{-0.15cm}=\hspace{-0.15cm}\left\{
\begin{aligned}
&\frac{1}{\sigma^2\gamma_\imath }\left|{\bar {\bm h}}_{\imath}^H	{\bar{\bm w}}_\imath \right|^2,  &\hspace{-1cm}\imath, \jmath \in {\cal{L}}~\text{and}~\imath=\jmath \\
&-\frac{1}{\sigma^2} \left|{\bar {\bm h}}_{\imath}^H	{\bar{\bm w}}_\jmath \right|^2,&\hspace{-1cm}\imath \in {\cal{L}}~\text{and}~\imath\ne \jmath  \\
&-{\bar{\bm w}}_\jmath^H  {\bm Q}_{L+1}  {\bar{\bm w}}_\jmath,&\hspace{-1cm}\imath=L+1~\text{and}~\jmath\in {\cal{L}} \\
&\frac{1}{\gamma_{L+1} } {\bar{\bm w}}_{L+1}^H  {\bm Q}_{L+1}  {\bar{\bm w}}_{L+1},  &\hspace{-1cm}\imath=\jmath=L+1 
\end{aligned}
\right..
\end{align}
In consequence, the transmit precoding power vector can be readily expressed as 
\begin{align}\label{power}
{\bm p}={\bar{\bm Q}}^{-1}{\bm 1}_{L+1}.
\end{align}
By combining \eqref{beamforming_j}, \eqref{beamforming_L}, and \eqref{power}, we obtain the structure of optimal transmit precoding as a function of the Lagrange multipliers $\{\lambda_j\}_{j=1}^{L+1}$.
The Lagrange multipliers can be effectively computed by solving dual problem (P3.3) with convex optimization (e.g., standard SDP or linear matrix inequalities (LMI) optimization) or leveraging the fixed-point equations \cite{Wiesel2006Linear,Yu2007Transmitter} which are given as \eqref{fixpeq} at the bottom of this page,
with ${\bar \lambda}_{\text{max}} \left[\cdot\right]$ denoting the largest eigenvalue of a matrix.

\subsubsection{ZF Transmit Precoding} Next, we consider applying the ZF criterion to eliminate
the multiuser interference among all low-mobility users, whose closed-form expression is given by
\setcounter{equation}{48} %
\begin{align}\label{ZF-based}
{\bar{\bm W}}_{ZF}={\bar{\bm H}}\left({\bar{\bm H}}^H {\bar{\bm H}}\right)^{-1} {\bar{\bm P}}
\end{align}
where ${\bar{\bm H}}^H=\left[{\bar {\bm h}}_{1},\ldots,{\bar {\bm h}}_{L}\right]^H$ denotes the equivalent channel matrix and
${\bar{\bm P}}=\text{diag}\left(\sqrt{P_1},\ldots,\sqrt{P_L}\right)$ denotes the diagonal transmit power matrix for the $L$ low-mobility users. Furthermore, we can set $P_l=\gamma_l\sigma^2$
to meet the constraints of \eqref{con1_P3} with equality so as to minimize the transmit power for each low-mobility user.

On the other hand, to avoid causing interference to low-mobility users, the transmit precoding ${\bm w}_{L+1}$ for high-mobility users can be designed by utilizing the null space of the channel matrix ${\bar{\bm H}}^H$. Specifically, we let ${\bm F} \in {\mathbb{C}^{M\times T}}$ with $M-L\le T\le M$ denote an orthonormal basis for the null space of ${\bar{\bm H}}^H$. By doing so, we can design the transmit precoding for high-mobility users as ${\bm w}_{L+1}={\bm F} {\bm v} $, with ${\bm v}\in {\mathbb{C}^{T\times 1}}$ being a linear combination vector. Apparently, we have ${\bar{\bm H}}^H{\bm w}_{L+1}={\bar{\bm H}}^H{\bm F} {\bm v}={\bm 0}_{L \times 1}$, which implies that the transmit precoding ${\bm w}_{L+1}$ for high-mobility users will not cause any interference to low-mobility users, i.e., ${\bm h}_{l}^H  {\bm w}_{L+1}=0, \forall l\in{\cal{L}}$ in \eqref{con1_P3}.
Thus, given the ZF transmit precoding in \eqref{ZF-based}, problem (P3) is reduced to
\begin{align}
\text{(P3.4):}
& \underset{ {\bm v}}{\text{min}}
& &   \left\|{\bm v}\right\|^2 \label{obj_P3.4} \\
& \hspace{-0.1cm} \text{s.t.} & & \hspace{-0.5cm}  \frac{  {\bm v}^H {\bm F}^H\left(  {\bm I}_M + 2 {\bm R}^H{\bm R} \right) {\bm F} {\bm v} }
{ {\rm tr} \hspace{-0.07cm} \left( \hspace{-0.07cm} {\bar{\bm W}}_{ZF}^H \hspace{-0.07cm} \left( {\bm I}_M \hspace{-0.1cm} + \hspace{-0.1cm} 2 {\bm R}^H{\bm R} \right) \hspace{-0.05cm}
	{\bar{\bm W}}_{ZF}\right) \hspace{-0.1cm} + \hspace{-0.1cm} {\bar \sigma}^2_{L+1} } \hspace{-0.1cm} \ge \hspace{-0.1cm} \gamma_{L+1} \label{con1_P3.4}
\end{align}
where $\left\|{\bm w}_{L+1}\right\|^2=\left\|{\bm F} {\bm v}\right\|^2={\bm v}^H {\bm F}^H {\bm F} {\bm v}={\bm v}^H{\bm v}=\left\|{\bm v}\right\|^2$ due to the orthonormality of ${\bm F}$, i.e., ${\bm F}^H {\bm F}={\bm I}_{T \times T}$.
Moreover, it can be readily demonstrated that the inequality constraint in \eqref{con1_P3.4} holds with equality at the optimal point; otherwise, one could scale down the optimal ${\bm v}$ to achieve equality with the constraint, thereby leading to a decrease in the objective function and thus contradicting the optimality.
Similar to \cite{Havary-Nassab2008Distributed}, it follows that the optimal ${\bm v}$ should be selected as the scaled principal eigenvector of ${\bm F}^H\left(  {\bm I}_M + 2 {\bm R}^H{\bm R} \right) {\bm F}$ associated with the largest eigenvalue, which is given by 
 \begin{align}\label{Null}
 {\bm v}^\star= \sqrt{p_{L+1}}{\bar{\bm v}} \quad \Rightarrow \quad  {\bm w}_{L+1}^\star={\bm F}{\bm v}^\star
 \end{align}
 where 
 ${\bar{\bm v}}={\mathbb P}\left\{ {\bm F}^H\left(  {\bm I}_M + 2 {\bm R}^H{\bm R} \right) {\bm F} \right\} $
 is the normalized principal eigenvector and thus ${\bm F}{\bar{\bm v}}$ is the normalized transmit precoding direction for high-mobility users, and $p_{L+1}$ denotes the precoding power, which is
 chosen to meet the equality of the constrain in \eqref{con1_P3.4}, i.e., 
 \begin{align}
 p_{L+1} \hspace{-0.05cm} = \hspace{-0.05cm} \frac{\gamma_{L+1} \hspace{-0.05cm} \left({\rm tr} \hspace{-0.05cm} \left({\bar{\bm W}}_{ZF}^H \left( \hspace{-0.05cm} {\bm I}_M \hspace{-0.05cm} + \hspace{-0.05cm} 2 {\bm R}^H{\bm R} \right) \hspace{-0.1cm}
 {\bar{\bm W}}_{ZF}\right) \hspace{-0.05cm} + \hspace{-0.05cm} {\bar \sigma}^2_{L+1}\right)}{  {\bar{\bm v}}^H {\bm F}^H\left(  {\bm I}_M + 2 {\bm R}^H{\bm R} \right) {\bm F} {\bar{\bm v}} }.
 \end{align}

\subsection{Reflect Precoding Optimization}\label{Reflect Beamforming}
For fixed transmit precoding ${\bm W}$ at the BS, problem (P2) reduces to a feasibility-check problem.
Specifically, by letting $a_{l,j}^*={\bm h}_{l}^H {\bm w}_j$ and ${\bm b}_{l,j}^H=e^{j\varphi}{\bm g}_{l}^H  {\rm diag}\left({\bm R}  {\bm w}_j\right) $,  problem (P2) is reduced to
\begin{align}
\text{(P4):}
& {\text{Find}}
& &  {\bar{\bm \theta}} \label{obj_P4} \\
& \text{s.t.} & & \frac{\left| {\bm b}_{l,l}^H {\bar{\bm \theta}} + a_{l,l}^* \right|^2}
{\sum\limits_{ j\ne l}^{L+1}\left|{\bm b}_{l,j}^H {\bar{\bm \theta}} + a_{l,j}^*\right|^2+\sigma^2}\ge \gamma_l, \forall l\in{\cal{L}} \label{con1_P4}\\
& & &|{\bar{\theta}}_{n}|=1, \forall n=1,\ldots,N. \label{con2_P4}
\end{align}
It can be readily verified that problem (P4) remains non-convex with its non-convex unit-modulus constraints in \eqref{con2_P4}.
Nevertheless, after transforming the constraints \eqref{con1_P4} and \eqref{con2_P4} into quadratic forms, we can efficiently solve problem (P4) by utilizing the SDR technique.

In specific, problem (P4) can be equivalently transformed into
\begin{align}
\text{(P4.1):}
& {\text{Find}}
& & {\bar{\bm \theta}} \label{obj_P4.1} \\
& \text{s.t.} & & {\tilde{\bm \theta}}^H \hspace{-0.05cm} {\bm B}_{l,l} {\tilde{\bm \theta}} + \left| a_{l,l} \right|^2 
\ge \nonumber \\
& & &\hspace{-0.3cm} \gamma_l \hspace{-0.05cm} \left( \hspace{-0.05cm} \sum_{ j\ne l}^{L+1} \hspace{-0.05cm} {\tilde{\bm \theta}}^H {\bm B}_{l,j} {\tilde{\bm \theta}} \hspace{-0.05cm} + \hspace{-0.13cm} \sum_{ j\ne l}^{L+1} \hspace{-0.1cm} \left|a_{l,j} \right|^2 \hspace{-0.1cm} + \hspace{-0.05cm} \sigma^2 \hspace{-0.1cm} \right) \hspace{-0.1cm}, \forall l \hspace{-0.05cm} \in \hspace{-0.05cm}{\cal{L}} \label{con1_P4.1}\\
& & &|t|=1, |{\bar{\theta}}_{n}|=1, \forall n=1,\ldots,N. \label{con2_P4.1}
\end{align}
where 
\begin{align}
{\bm B}_{l,j}=\begin{bmatrix}
{\bm b}_{l,j} {\bm b}_{l,j}^H&a_{l,j}^*{\bm b}_{l,j} \\a_{l,j} {\bm b}_{l,j}^H&0
\end{bmatrix}, \quad
{\tilde{\bm \theta}}=\begin{bmatrix}{\bar{\bm \theta}} \\t \end{bmatrix}
\end{align}
with $t$ being an auxiliary variable.
For the ease of calculation, we further define ${\tilde{\bm \Theta}}={\tilde{\bm \theta}} {\tilde{\bm \theta}}^H$ by exploiting the property of ${\tilde{\bm \theta}}^H {\bm B}_{l,j} {\tilde{\bm \theta}} =\text{tr} \left({\bm B}_{l,j} {\tilde{\bm \theta}} {\tilde{\bm \theta}}^H \right)$. Note that ${\tilde{\bm \Theta}}$ is required to satisfy the conditions of ${\tilde{\bm \Theta}} \succeq {\bm 0}$ and $\text{rank}\left({\tilde{\bm \Theta}}\right)=1$. 
However, the rank-one constraint is non-convex and solving the feasibility-check problem may not lead to an efficient reflect precoding for reducing the total transmit power at the BS.
To obtain a more efficient solution, we relax the rank-one constraint and reformulate problem (P4.1) with an explicit objective, which is given by
\begin{align}
\text{(P4.2):}
& \underset{ {\tilde{\bm \Theta}}, \left\{c_l\right\}_{l=1}^L}{\text{max}}
& & \sum_{l=1}^{L} c_l \label{obj_P4.2} \\
& \hspace{0.1cm} \text{s.t.} & & \hspace{-0.7cm} \text{tr} \left( {\bm B}_{l,l} {\tilde{\bm \Theta}} \right) + \left|a_{l,l}\right|^2 \ge \nonumber \\
& & & \hspace{-1cm} \gamma_l  \hspace{-0.1cm} \left( \hspace{-0.1cm}  \sum_{ j\ne l}^{L+1} \hspace{-0.1cm} \text{tr} \hspace{-0.08cm} \left( \hspace{-0.05cm} {\bm B}_{l,j} {\tilde{\bm \Theta}} \hspace{-0.1cm} \right) \hspace{-0.1cm} + \hspace{-0.12cm} \sum_{ j\ne l}^{L+1} \hspace{-0.1cm} \left| \hspace{-0.02cm} a_{l,j} \hspace{-0.05cm} \right|^2  \hspace{-0.17cm} + \hspace{-0.05cm} \sigma^2 \hspace{-0.08cm} \right) \hspace{-0.12cm} + \hspace{-0.05cm} c_l, \hspace{-0.05cm} \forall l \hspace{-0.05cm} \in \hspace{-0.05cm} {\cal{L}}\hspace{-0.05cm} \label{con1_P4.2}\\
& & & \hspace{-0.6cm} \left[{\tilde{\bm \Theta}}\right]_{n,n}=1,\quad \forall n=1,\ldots,N+1 \label{con2_P4.2}\\
& & & \hspace{-0.5cm} {\tilde{\bm \Theta}} \succeq {\bm 0}, c_l\ge 0, \quad \forall l\in{\cal{L}}
\end{align}
where the slack variable $c_l$ can be considered as the ``SINR residual" of low-mobility user $l$ in precoding optimization.
Note that problem (P4.2) is a convex SDP problem, and thus we can optimally address it using established convex optimization solvers including CVX \cite{grant2014cvx}.
Since such SDR technique might not yield a rank-one solution, we can still recover a high-quality rank-one solution for problem (P4.1) by employing methods like Gaussian randomization based on the obtained higher-rank solution \cite{Wu2019TWC}.

\subsection{Convergence Performance and Complexity}\label{Convergence Performance}
\begin{algorithm}[t]
	\caption{Proposed AO Algorithm for Solving (P1)} \label{alg1}
	\begin{algorithmic}[1]
		
		\STATE Initialization: Initialize the reflection precoding vector ${\bar{\bm \theta}}:={\bar{\bm \theta}}^{(0)}$, the transmit precoding matrix ${\bm W}:={\bm W}^{(0)}$, and the iteration number $i=0$.
		
		\REPEAT
		
		
		\STATE Solve (P3) with given ${\bar{\bm \theta}}^{(i)}$ through the MMSE/ZF transmit precoding design in Section~\ref{Transmit Beamforming}, and represent the solution as ${\bm W}^{(i)}$.
		
		\STATE Solve (P4.2) with given ${\bm W}^{(i)}$ through the SDR method in Section~\ref{Reflect Beamforming}, and set the solution after applying Gaussian randomization as ${\bar{\bm \theta}}^{(i+1)}$.
		
		\STATE Update $i:=i+1$.
		
		\UNTIL {The increment of the objective value in \eqref{obj_P2} is less than a pre-designed threshold $\xi>0$ or the iteration number $i$ is larger than a preset number of iterations $I$.}
	\end{algorithmic}
\end{algorithm}
We address problem (P2) in the proposed AO algorithm by alternately solving sub-problems (P3) and (P4.2) in an iterative manner, where the solution of each iteration serves as the initial point for the subsequent iteration.
In Algorithm~\ref{alg1}, we summarize the procedures of our proposed algorithm for solving problem (P2) to minimize the total transmit power at the BS.

\subsubsection{Convergence Performance}
Intuitively, if solving (P4.2) results in a feasible solution with an SINR that is strictly larger than the SINR target $\gamma_l$ for $l \in \mathcal{L}$, then one can properly reduce both the transmit power of user $l$ and the total transmit power in problem (P3) without violating any of the SINR constraints.
As such, the guarantee of convergence for the proposed AO algorithm stems from the fact that the objective value of (P3) remains non-increasing during the iterations. Additionally, the optimal objective value of (P3) is lower-bounded by a finite value, such as $0$.

\subsubsection{Complexity} \rev{It can be easily verified that the complexity of solving problem (P4.2) via the SDR method is ${\cal O}(N^{4.5})$ \cite{Luo2010Semidefinite}.
Furthermore, the MMSE/ZF precoding design for addressing problem (P3), which involves matrix inversion using methods such as the Gauss-Jordan elimination, results in a computational complexity of ${\cal O}(M^3)$.
Hence, the overall complexity of Algorithm~\ref{alg1} is ${\cal O}\left(I( N^{4.5} +M^3)\right)$, where $I$ represents the number of iterations required for convergence.}

\section{Simulation Results}\label{Sim}
\begin{figure}[!t]
	\centering
	\includegraphics[width=3.5in]{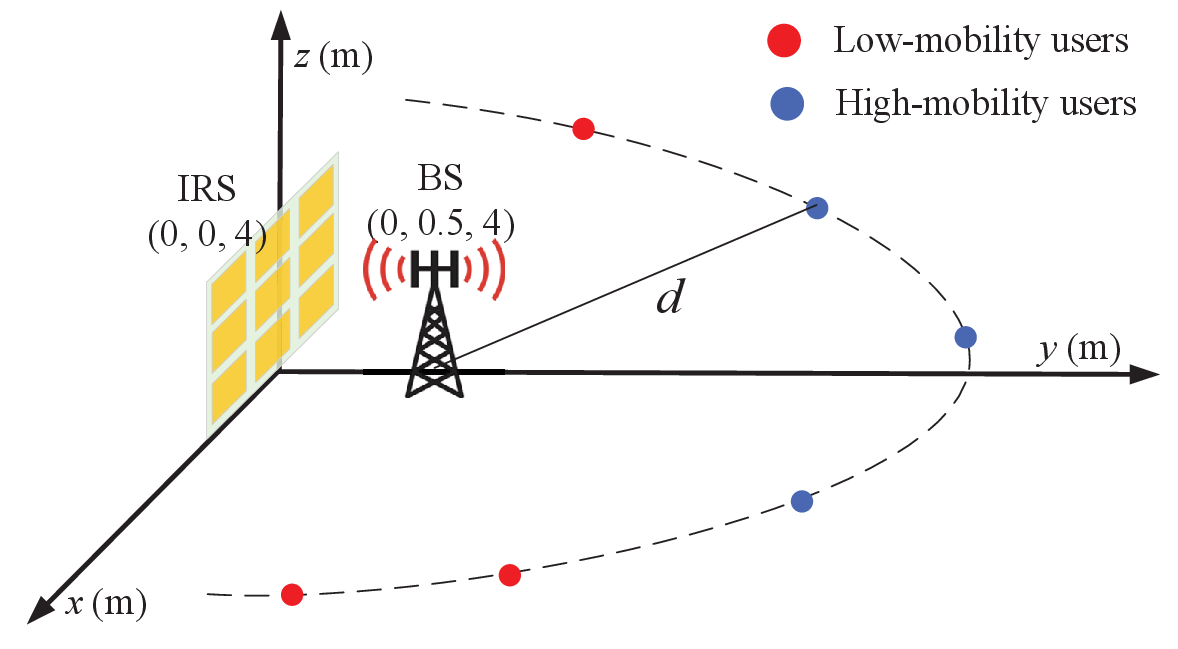}
	\setlength{\abovecaptionskip}{-6pt}
	\caption{Simulation setup for IRS-aided multiuser downlink communication.}
	\label{system_sim}
\end{figure}
In this section, simulation results of our proposed co-design of transmit diversity and active/passive precoding in an IRS-aided multiuser communication system for simultaneously serving low-mobility and high-mobility users are provided.
The simulated system configuration is illustrated in Fig.~\ref{system_sim}, where the IRS and the multi-antenna BS are located in one three-dimensional (3D) Cartesian coordinate system with their center points at $(0,0,4)$ m and $(0,0.5,4)$ m, respectively.
Specifically, the IRS and the BS are placed on the $x-z$ plane and $y-z$ plane, respectively.
Without loss of generality, we assume that all users, regardless of their mobility, are randomly situated within the front half-space reflection area of the IRS with identical distance of $d=d_j=50$ m, $j\in{\cal{L}}\cup{\cal{K}}$ on the $x-y$ plane, and both the numbers of low-mobility and high-mobility users are set as $3$, i.e., totally 6 users in the system. In addition, it is assumed that all the low-mobility users share an identical SINR target, i.e., $\gamma=\gamma_l, \forall l\in{\cal{L}}$ and all the high-mobility users also have the same SINR target of $\gamma_{L+1}$ for their common information.
 
We utilize the square uniform planar array (UPA) model for the IRS, where the quantity of reflecting elements along both the $x$- and $z$-axes is equal, i.e., $N_x=N_z=\sqrt{N}$.
To simplify the implementation in practical IRS-assisted multiuser communication systems, we group every $5\times 5$ adjacent IRS elements, which share the same phase shift, into a subsurface \cite{zheng2019intelligent}.
For all individual links, we set the reference path gain at a distance of $1$ m as $\beta=-30$~dB.
The path loss exponent is set as $\alpha=2$ for the link between the BS and its integrated IRS (which is modeled by the near-field line-of-sight (LoS) channel due to the short distance \cite{Zheng2022Simultaneous}) and set as $\alpha=2.5$ for other links (due to the relatively large distance).
In addition, the channels between the BS/IRS and the low-mobility users, i.e., $\left\{{\bm g}_{l}^H, {\bm h}_{l}^H\right\}, \forall l\in{\cal{L}}$, are modeled as the Rician fading with a       Rician factor of $5$ dB.
Unless otherwise stated, the number of BS antennas is $M=8$,
the wavelength is $\lambda=0.05$~m, the noise power at each user is set equal as $\sigma^2=-85$~dBm, and the element spacing is set as $\Delta=\lambda/2=0.025$~m.

\begin{figure}[!t]
	\centering
	\includegraphics[width=3.5in ]{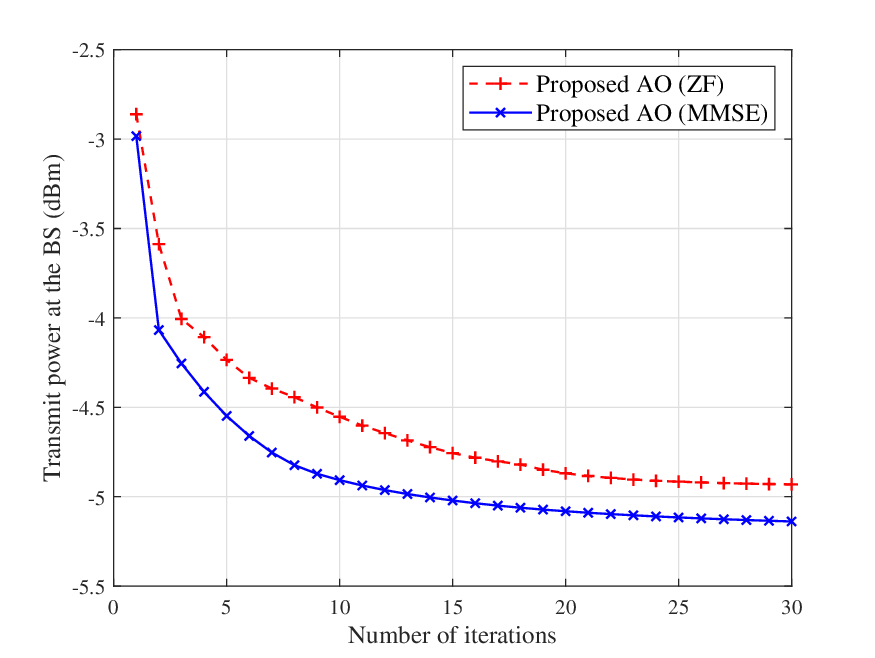}
	\setlength{\abovecaptionskip}{-5pt}
	\caption{\rev{Convergence behavior of Algorithm 1, with $N=400$, $\gamma=8$, and $\gamma_{L+1}=1$.}}
	\label{Convergence}
\end{figure}
\subsection{Active/Passive Precoding}
\rev{For the collaborative design of active and passive precoding in problem (P2), we consider two benchmark schemes as follows.
\begin{itemize}
	\item {\bf Random phase with MMSE transmit precoding}: In this scheme, the reflect coefficients in ${\bar{\bm \theta}}$ are randomly drawn from a uniform distribution within $[0, 2\pi)$ at the IRS and then the MMSE transmit precoding in Section~\ref{Transmit Beamforming} is applied at the BS.
	\item {\bf DFT-based codebook search with MMSE transmit precoding}: In this scheme, the passive precoding ${\bar{\bm \theta}}$ is searched over a discrete Fourier transform (DFT)-based codebook (denoted by ${\cal D}$) to maximize the minimum channel gain among all low-mobility users, i.e., ${\bar{\bm \theta}}^\star=\arg \underset{ {\bar{\bm \theta}}\in{\cal D} }{\text{max}}~\underset{ {\bm l\in{\cal{L}}}}{\text{min}} \|{\bar {\bm h}}_{l}^H\|^2$. Then, given ${\bar{\bm \theta}}^\star$, the MMSE transmit precoding in Section~\ref{Transmit Beamforming} is applied at the BS.
\end{itemize}}

\rev{Prior to comparing the performance of the proposed algorithm with the benchmark schemes, we first demonstrate the convergence behavior of Algorithm~\ref{alg1} in Fig.~\ref{Convergence}
by setting $N=400$, $\gamma=8$, and $\gamma_{L+1}=1$. The (passive) precoding vector is initialized as ${\bar{\bm \theta}}={\bm 1}_{N\times 1}$. We observe that the transmit power required by the proposed AO algorithm decreases both monotonically and rapidly with the number of iterations, which validates the convergence performance analysis presented in Section~\ref{Convergence Performance}.
Furthermore, applying MMSE transmit precoding in Algorithm 1, as opposed to ZF transmit precoding, results in a lower converged power.}

\begin{figure}[!t]
	\centering
	\includegraphics[width=3.5in ]{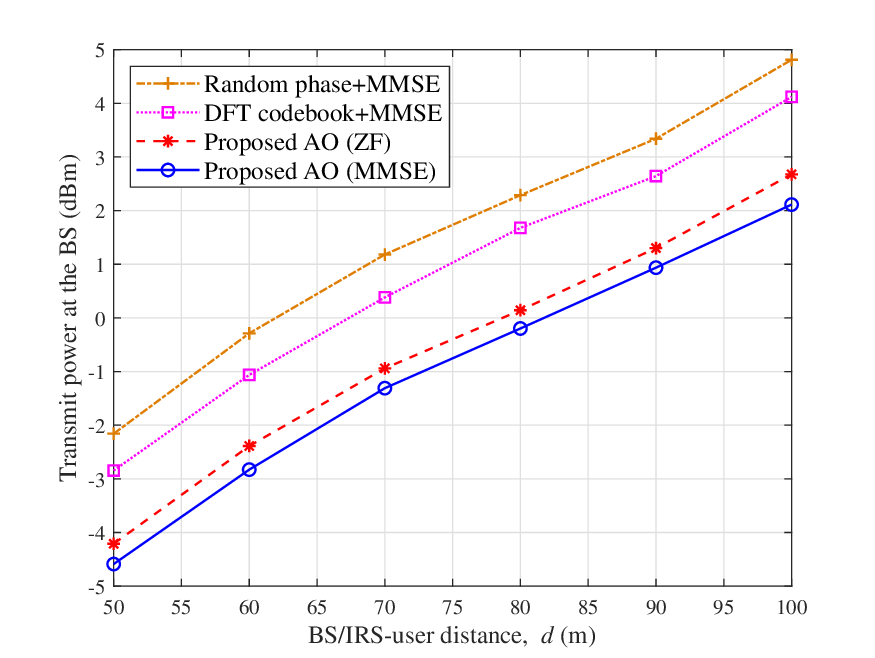}
	\setlength{\abovecaptionskip}{-5pt}
	\caption{BS transmit power versus BS/IRS-user distance, with $N=400$, $\gamma=10$, and $\gamma_{L+1}=1$.}
	\label{Power_vs_DISTalg}
\end{figure}

In Fig.~\ref{Power_vs_DISTalg}, we compare the transmit power at the BS with regard to BS/IRS-user distance for different designs in the IRS-aided multiuser communication system, with $N=400$, $\gamma=10$, and $\gamma_{L+1}=1$. It is observed that regardless of the BS/IRS-user distance, the proposed AO algorithm with the MMSE transmit precoding achieves the lowest transmit power and also outperforms its ZF counterpart. In particular, owing to the near-optimal passive precoding design, the proposed AO algorithm with the MMSE transmit precoding requires about $2$~dB and $2.7$~dB lower transmit power than the DFT-based codebook search and the random phase schemes, respectively.


\begin{figure}[!t]
	\centering
	\includegraphics[width=3.5in ]{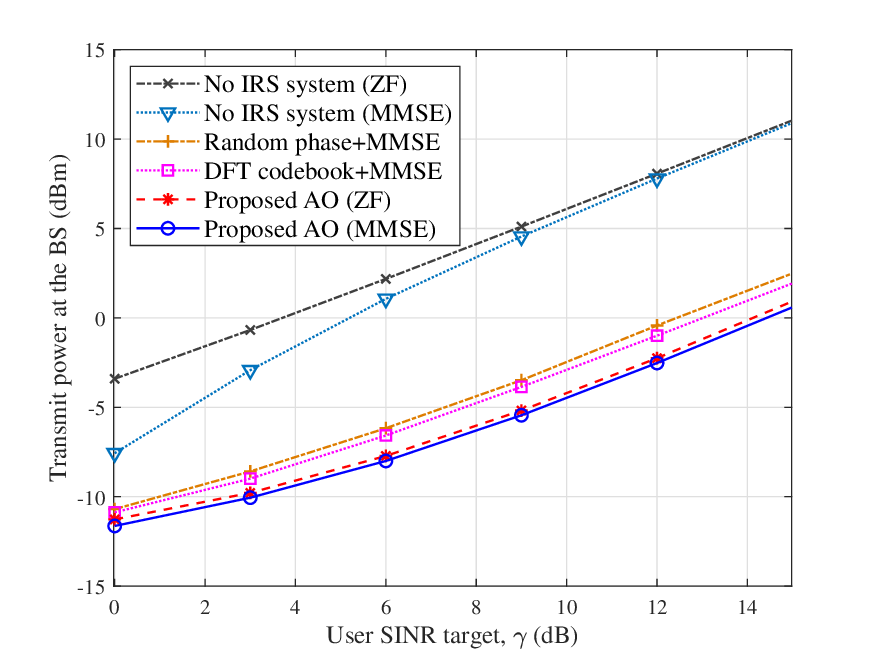}
	\setlength{\abovecaptionskip}{-5pt}
	\caption{\rev{BS transmit power versus the SINR target of low-mobility users, with $N=400$ and $\gamma_{L+1}=1$.}}
	\label{Power_vs_USINRsys}
\end{figure}

Next, to demonstrate the performance gain brought by IRS,
we consider a multiuser MISO communication without IRS as a baseline system. 
Specifically, the MMSE and ZF transmit precoding designs developed in Section \ref{Transmit Beamforming} can be applied to the multiuser MISO communication without IRS, by setting ${\bar{\bm \theta}}={\bm 0}_{L \times 1}$ and ${\bm R}={\bm 0}_{N \times M}$.\footnote{For high-mobility users, we can design two orthogonal transmit precoding vectors. These vectors create an orthogonal channel condition for the transmission of information ${\tilde s}_1$ and ${\tilde s}_2$. This design achieves transmit diversity and emulates the space-time code found in the classic Alamouti's scheme (see Table \ref{coding}).}
\rev{In Fig. \ref{Power_vs_USINRsys}, we compare the transmit power at the BS versus the SINR target of low-mobility users under multiuser MISO systems with and without IRS.
One can observe that the IRS-aided system using the proposed AO algorithm requires significantly lower transmit power (up to $10.3$ dB) than the baseline system without IRS, thanks to the pronounced passive precoding gain.
Moreover, as the SINR target increases, a tiny gap (about $0.3$ dB) exists between the MMSE and ZF transmit precoding in the proposed IRS-aided system, which is due to the coupling effect between the transmit and reflect precoding. In contrast, the MMSE transmit precoding asymptotically approaches the ZF one in the baseline system without IRS due to the reduced noise effect as the SINR target increases.}

\begin{figure}[!t]
	\centering
	\includegraphics[width=3.5in ]{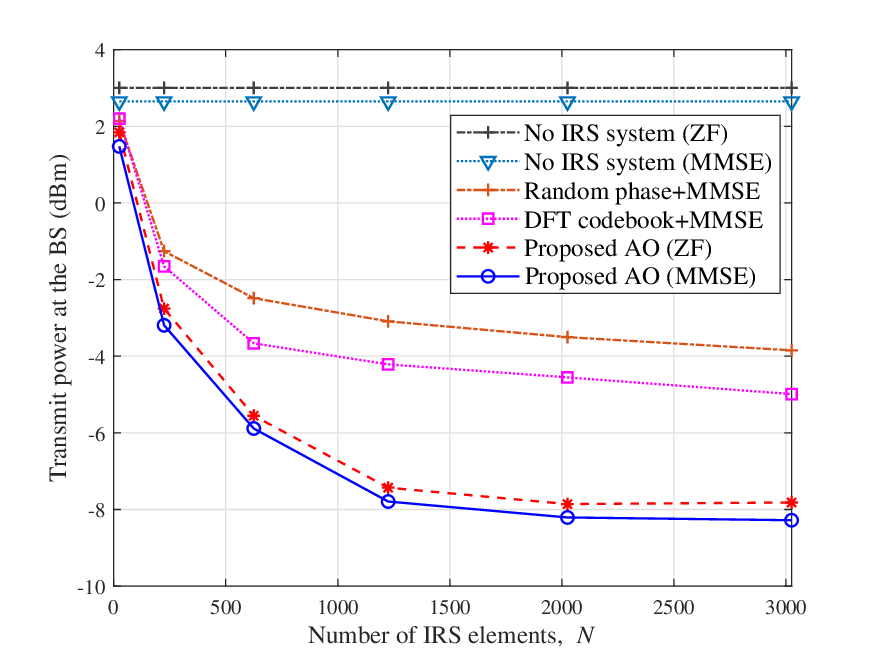}
	\setlength{\abovecaptionskip}{-5pt}
	\caption{\rev{BS transmit power versus the number of IRS elements, with $\gamma=10$ and $\gamma_{L+1}=1$.}}
	\label{Power_vs_numIRS}
\end{figure}

In Fig.~\ref{Power_vs_numIRS}, we illustrate the transmit power at the BS versus the number of IRS elements for both systems with and without IRS.
\rev{First, the transmit power at the IRS-integrated BS decreases with the increase of IRS element number $N$, owing to the increased signal reflection power for achieving higher energy efficiency.
Second, regardless of $N$, the proposed AO algorithm achieves the lowest transmit power and significantly outperforms the two benchmark schemes as well as the baseline system without IRS.
Nevertheless, one can find that when $N\ge 2000$, the transmit power reduction of the proposed AO algorithm becomes marginal with the further increase of the IRS element number $N$.
This is attributed to the fact that under the near-field LoS channel model between the co-located IRS and BS, the passive precoding gain exhibits a diminishing return and eventually reaches a limit.}

\subsection{Transmit Diversity}

\rev{Next, we further demonstrate the symbol error rate (SER) performance of our proposed IRS-aided transmit diversity scheme,
where the $8$-ary PSK modulation and the following benchmark schemes are considered. The following results are obtained via Monte Carlo simulations.}
\begin{itemize}
	\item {\bf Null-space based scheme without IRS}: In this scheme, ${\bm w}_{L+1}$ is designed to project into the null space of 
	${\bar{\bm H}}^H$ and then the BS multicasts the modulated symbols ${\bm w}_{L+1}{\tilde s}$ (without space-time code design) to high-mobility users (with no interference to low-mobility users).
	\item {\bf Null-space based scheme with dumb IRS}: In this scheme, the IRS fixes its common phase-shift and the BS multicasts the modulated symbols ${\bm w}_{L+1}{\tilde s}$ (without space-time code design) to high-mobility users with ${\bm w}_{L+1}$ projected into the null space of ${\bar{\bm H}}^H$. 
	\item {\bf Null-space based Alamouti's scheme}: In this scheme, 
	two orthogonal transmit precoding vectors ${\bm w}_{L+1}$ and ${\bm w}'_{L+1}$ (based on the null space of ${\bar{\bm H}}^H$) are generated to
	create an orthogonal channel condition for transmitting the space-time code as in Table~\ref{coding} to high-mobility users.
\end{itemize}

\begin{figure}[!t]
	\centering
	\includegraphics[width=3.5in ]{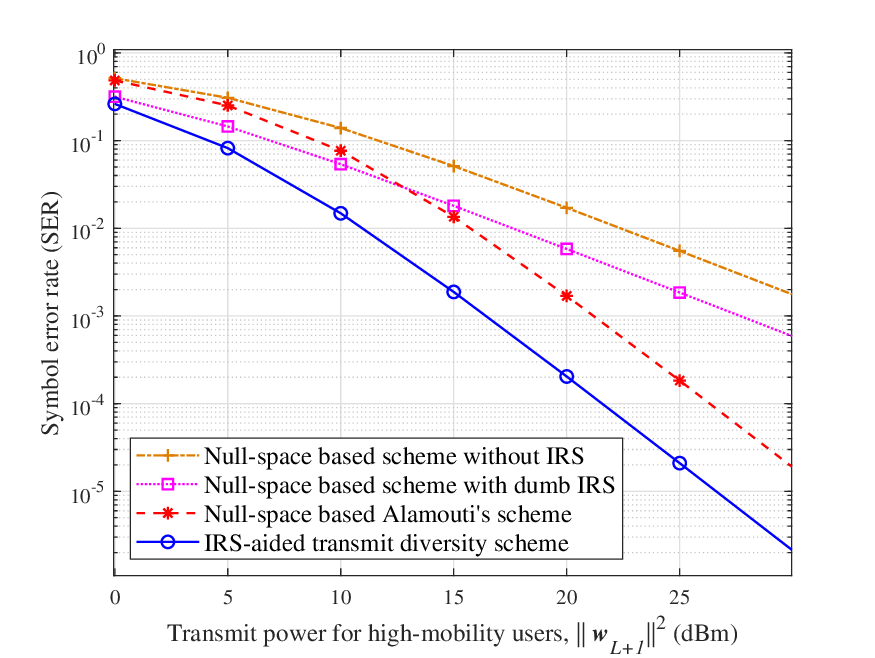}
	\setlength{\abovecaptionskip}{-5pt}
	\caption{SER versus BS transmit power for high-mobility users, with $N=400$ and $\gamma=10$.}
	\label{SER_vs_power}
\end{figure}
In Fig.~\ref{SER_vs_power}, we exhibit the average SER with regard to the transmit power at the BS for high-mobility users, with $N=400$ and $\gamma=10$.
The following are some interesting observations of this figure.
First, as compared to the null-space based scheme without IRS and with dumb IRS, both of which lack transmit diversity, our proposed IRS-aided transmit diversity scheme and the null-space based Alamouti’s scheme manage to achieve a transmit diversity gain of order two. This is accomplished by creating an orthogonal channel condition, thereby dramatically enhancing the SER performance.
Second, the IRS-aided transmit diversity scheme (null-space based scheme with dumb IRS) attains up to a $5$ dB gain over the null-space based Alamouti’s scheme (null-space based scheme without IRS). This gain is due to the increased signal reflection power introduced by the IRS, leading to a higher average channel gain. 
Third, at low-to-medium power levels (i.e., $\left\|{\bm w}_{L+1}\right\|^2\le 12$ dBm), the null-space based scheme with dumb IRS even surpasses the null-space based Alamouti’s scheme. Despite the transmit diversity, the null-space based scheme with dumb IRS can achieve a higher average channel gain due to the additional IRS-reflected channel.

\begin{figure}[!t]
	\centering
	\includegraphics[width=3.5in ]{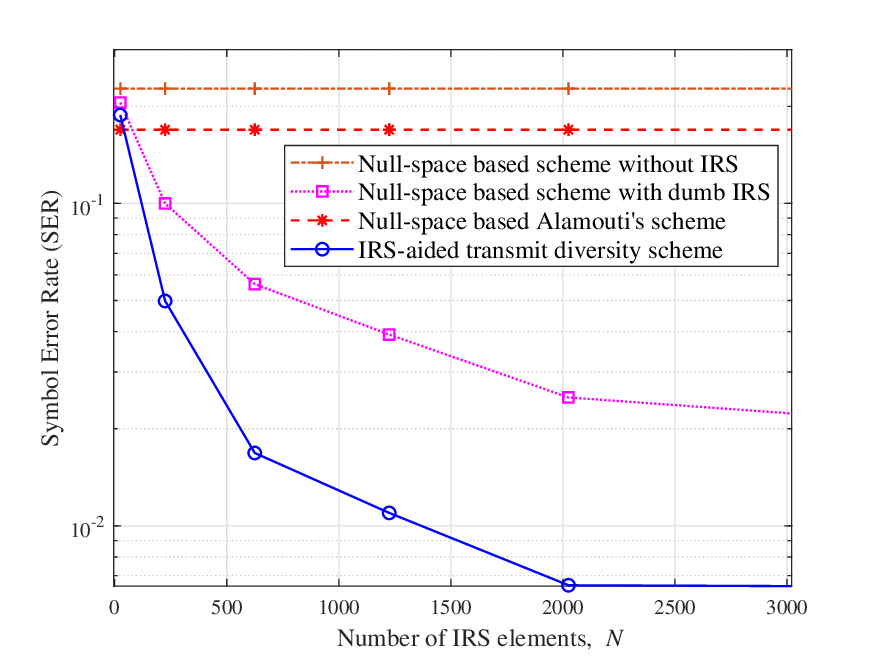}
	\setlength{\abovecaptionskip}{-5pt}
	\caption{SER versus  the number of IRS elements, with $\gamma=10$ and $\left\|{\bm w}_{L+1}\right\|^2=10$~dBm.}
	\label{SER_vs_numIRS}
\end{figure}
 
In Fig.~\ref{SER_vs_numIRS}, we depict the SER in relation to the number of IRS elements for various schemes, with $\gamma=10$ and $\left\|{\bm w}_{L+1}\right\|^2=10$~dBm.
The figure leads to several observations.
First, the SERs of both the proposed scheme and the null-space based scheme with dumb IRS decrease as ${\bar N}$ increases. This is due to the additional signal reflection power in the IRS-reflected link.
Second, by leveraging the transmit diversity gain provided by the IRS, our proposed scheme attains the lowest SER among all the schemes.
Third, a SER floor is observed in both the proposed scheme and the null-space based scheme with dumb IRS when $N\ge 2000$.
This suggests that the IRS reflection power increases with the number of reflecting elements until it reaches a constant value.

\section{Conclusions}\label{conlusion}
In this paper, we explored an innovative IRS-aided multiuser communication system with a new architecture of IRS-integrated BS, and proposed a co-design of transmit diversity and active/passive precoding to simultaneously serve multiple high-mobility and low-mobility users without and with CSI, respectively.
Specifically, we utilized the common phase-shift of the IRS to serve high-mobility users via transmit diversity at the BS, eliminating the need for CSI. Concurrently, the active/passive precoding gain can be achieved at the IRS-integrated BS to serve low-mobility users, provided their CSI is known.
We then formulated and solved a joint reflect and transmit precoding optimization problem to minimize the total transmit power at the BS.
Simulation results validated the superior performance gain achieved by the IRS-aided multiuser communication system with our proposed co-design of transmit diversity and active/passive precoding in various system settings.



\ifCLASSOPTIONcaptionsoff
  \newpage
\fi
\bibliographystyle{IEEEtran}
\bibliography{AlamoutiIRS}

\end{document}